\documentclass[prb,reprint,superscriptaddress,a4paper,aps,showpacs]{revtex4-1}
\usepackage{graphicx}

\begin{document}

\newcommand{\ybsn}{Yb$_2$Sn$_2$O$_7$}
\newcommand{\ybti}{Yb$_2$Ti$_2$O$_7$}
\newcommand{\tbsn}{Tb$_2$Sn$_2$O$_7$}
\newcommand{\tbti}{Tb$_2$Ti$_2$O$_7$}
\newcommand{\m}{$\mu_{\rm B}$}

\title{Evidence for unidimensional low-energy excitations as the origin of 
persistent spin dynamics in geometrically frustrated magnets}

\author{A.~Yaouanc}
\affiliation{Universit\'e Grenoble Alpes, INAC-SPSMS, F-38000 Grenoble, France}
\affiliation{CEA, INAC-SPSMS, F-38000 Grenoble, France}
\affiliation{Laboratory for Muon-Spin Spectroscopy, Paul Scherrer Institute,
CH-5232 Villigen-PSI, Switzerland}
\author{P.~Dalmas~de~R\'eotier}
\affiliation{Universit\'e Grenoble Alpes, INAC-SPSMS, F-38000 Grenoble, France}
\affiliation{CEA, INAC-SPSMS, F-38000 Grenoble, France}
\author{A.~Bertin}
\affiliation{Universit\'e Grenoble Alpes, INAC-SPSMS, F-38000 Grenoble, France}
\affiliation{CEA, INAC-SPSMS, F-38000 Grenoble, France}
\author{C.~Marin}
\affiliation{Universit\'e Grenoble Alpes, INAC-SPSMS, F-38000 Grenoble, France}
\affiliation{CEA, INAC-SPSMS, F-38000 Grenoble, France}
\author{E.~Lhotel}
\affiliation{Institut N\'eel, CNRS and Universit\'e Joseph Fourier, 
BP 166, F-38042 Grenoble Cedex 9, France}
\author{A.~Amato}
\affiliation{Laboratory for Muon-Spin Spectroscopy, Paul Scherrer Institute,
CH-5232 Villigen-PSI, Switzerland}
\author{C.~Baines}
\affiliation{Laboratory for Muon-Spin Spectroscopy, Paul Scherrer Institute,
CH-5232 Villigen-PSI, Switzerland}

\date{\today}

\begin{abstract}

We report specific heat, magnetic, and muon spin relaxation measurements performed on a polycrystalline sample of the normal spinel CdHo$_2$S$_4$. The rare-earth ions sit on a lattice of corner-sharing regular tetrahedra as in pyrochlore compounds. Magnetic ordering is detected at $T_{\rm c} \simeq 0.87$~K. From spin-lattice relaxation rate measurements on both sides of $T_{\rm c}$ we uncover similar magnetic excitation  modes driving the so-called persistent spin dynamics at $T < T_{\rm c}$. Unidimensional excitations are argued to be at its origin. Often observed spin loop structures are suggested to support these excitations. The possibility of a generic mechanism for their existence is discussed.

\end{abstract}


\maketitle
 
Magnetic materials with coupled spins located on corner-sharing tetrahedra are 
expected to exhibit geometrical magnetic frustration because their spatial arrangements 
are such that they prevent the simultaneous minimization of all the interaction 
energies. Typical examples are given by the pyrochlore insulator compounds of generic 
chemical formula $R_2M_2$O$_7$, where $R$ is a rare earth ion and $M$ a non magnetic 
element.\cite{Gardner10} For instance Ho$_2$Ti$_2$O$_7$ 
for which the net interaction between the spins is ferromagnetic, has been the first 
recognized spin-ice system and an analogy has been drawn between the proton positions in common ice I$_{\rm h}$ 
and the spin configuration.\cite{Harris97} Its properties seem mostly described by classical 
physics. On the other hand, Yb$_2$Ti$_2$O$_7$ is an 
example of a three-dimensional quantum spin liquid,\cite{Hodges02} at least for the
best available sample characterized by a clear specific heat peak at its
first order transition.\cite{Yaouanc11c,Ross11a} While this transition is reminiscent of that observed between the gas and liquid states of conventional matter, it is characterized by a strong quantum entanglement.\cite{Balents10}
As a last example, we cite Yb$_2$Sn$_2$O$_7$ which is a splayed ferromagnet, i.e.\ essentially a ferromagnetic compound,\cite{Yaouanc13} with an emergent gauge field.\cite{Savary12}

Although this physics is exotic, an interpretation at the mean-field level is within reach.\cite{Savary12}
However, the most exotic property of these compounds lies in their dynamics. 
The origin of the ubiquitous persistent spin dynamics observed in 
geometrically frustrated magnetic materials is still elusive. Its most famous 
signature is a finite and approximately temperature independent spin-lattice relaxation rate $\lambda_Z$ observed in these compounds below $\approx 1$~K irrespective of the presence of a magnetic order or not. Conventionally, at temperatures well below $|\theta_{\rm CW}|$ where $\theta_{\rm CW}$ is the Curie-Weiss temperature, a magnetic system should order and  $\lambda_Z$ should vanish when the temperature approaches zero. Still, a finite and temperature independent $\lambda_Z$ is found in the ordered state of magnetic compounds such as Cu$_2$Cl(OH)$_3$,\cite{Zheng05} Gd$_2$Sn$_2$O$_7$,\cite{Dalmas04,Chapuis09b} and Gd$_2$Ti$_2$O$_7$,\cite{Yaouanc05a} to cite few reported cases. Also surprising is the absence of muon spin spontaneous precession in muon spin relaxation ($\mu$SR) measurements below the magnetic 
critical temperature $T_{\rm c}$
for Tb$_2$Sn$_2$O$_7$,\cite{Dalmas06,Bert06} Er$_2$Ti$_2$O$_7$,\cite{Lago05,Dalmas12a} and Yb$_2$Sn$_2$O$_7$.\cite{Yaouanc13} This lack of spontaneous precession stems from unexpected excitation modes below $T_{\rm c}$, as first shown for Tb$_2$Sn$_2$O$_7$.\cite{Dalmas06,Chapuis07,Rule09b} A physical mechanism accounting for them is still missing.

\begin{figure}
\centering
\begin{picture}(245,150)
\put(0,20){
\includegraphics[width=0.225\textwidth]{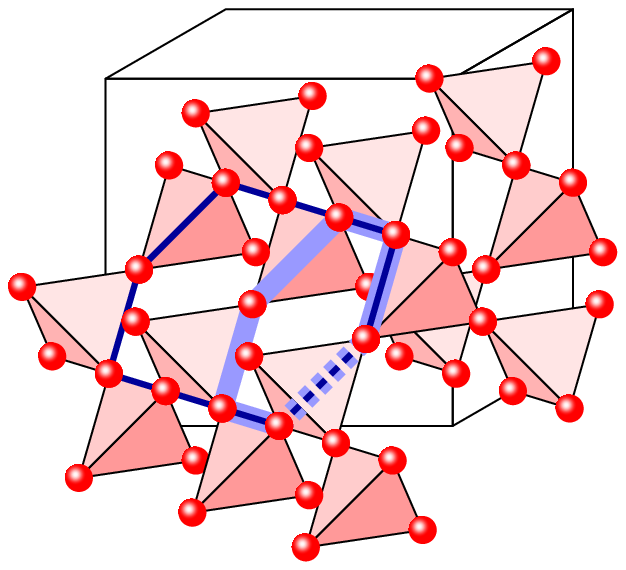}
}
\put(128,0){
\includegraphics[width=0.225 \textwidth]{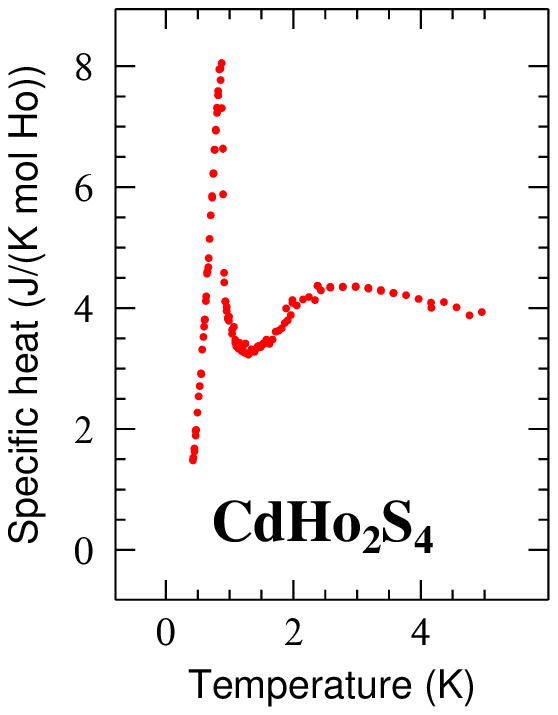}
}
\put(0,127){(a)}
\put(224,127){(b)}
\end{picture}
\caption{(Color online)
(a) Rare-earth ions lattice in the pyrochlore $R_2M_2$O$_7$ and normal spinel Cd$R_2X_4$ compounds. The thicker light blue (thinner dark blue) bold line represents a 6 (10)-site loop. (b) Low temperature heat capacity of CdHo$_2$S$_4$. 
}
\label{Structure_specific_heat_data}
\end{figure}

Recently normal spinels of chemical formula Cd$R_2X_4$ ($X$ = S, Se) have attracted some attention.\cite{Lau05} In this crystal structure, the $R$ ions form the same lattice of corner-sharing regular tetrahedra as in the pyrochlore compounds; see Fig.~\ref{Structure_specific_heat_data}a. A spin-ice behavior has been discovered for CdEr$_2$Se$_4$,\cite{Lago10} highlighting the interest of extending the number of compounds with geometrical frustration on a three-dimensional lattice.

Here we report bulk and $\mu$SR measurements for the normal thiospinel CdHo$_2$S$_4$. We find evidence for a magnetic phase transition with similar magnetic excitation modes on both sides of $T_{\rm c}$. We show that these modes are at the origin of the observed persistent spin dynamics as fingerprinted by $\lambda_Z$.

The synthesis of CdHo$_2$S$_4$ powder has followed a two step route. First, Ho$_2$S$_3$ has been prepared starting with holmium metal (4N) and sulfur (5N) properly mixed and heat treated in a vacuum sealed quartz tube up to 720$^\circ$C over 2 weeks, the temperature being increased step by step to avoid excessive pressure due to sulfur vapor. The phase has been checked by x-ray powder diffraction. Secondly, Ho$_2$S$_3$ has been mixed with commercial CdS powder (5N) and pressed under 4 tons into 13 mm diameter pellets to improve solid state reaction. A heat treatment of the resulting product has been achieved up to 900$^\circ$C for two weeks in a sealed quartz tube under vacuum. The final yellow/brown ceramic has been found to be the CdHo$_2$S$_4$ phase perfectly crystallized without any x-ray detected foreign phases. Finally sintered pellets were obtained from this ceramic after grinding and compaction with the same press, followed by a heat treatment at 600$^\circ$C over 6 hours under vacuum. 

The investigation of the macroscopic properties has consisted of measurements of the heat capacity using a Physical Property Measurement System (Quantum Design Inc.), and of the magnetization and ac susceptibility. These magnetization experiments have been performed by the extraction method using a Magnetic Property Measurement System (Quantum Design Inc.) for measurements down to 2~K and a superconducting quantum interference device magnetometer developed at the Institut N\'eel\cite{Paulsen01} for measurements down to 0.07~K and up to an external magnetic field $B_{\rm ext} =8$~T.

The $\mu$SR experiments were carried out at the Swiss Muon Source (S$\mu$S, Paul Scherrer Institute, Switzerland) either at the Low Temperature Facility (LTF) or the General Purpose Surface-muon instrument (GPS) depending of the temperature range. Measurements were performed with the transverse (longitudinal) geometry in which the external field defining the $Z$ axis of a referential frame, is applied perpendicular (parallel) to the initial muon spin polarization. The measured physical quantity is the so-called $\mu$SR asymmetry time spectrum which describes the evolution of the projection of the muon polarization perpendicular to (along) the direction of the initial polarization.\cite{Yaouanc11} The spectrum is denoted as $a_0 P^{\rm exp}_X(t)$ ($a_0 P^{\rm exp}_Z(t)$) for the transverse (longitudinal) field geometry. Zero applied field measurements were also performed in the longitudinal geometry.

The heat capacity $C_{\rm p}$ depicted in Fig.~\ref{Structure_specific_heat_data}b displays a fairly narrow peak at $T_{\rm c} \simeq 0.86$~K, signalling a thermodynamic phase transition, and a broad hump centered at about 3~K. This latter feature, attributed to the onset of short-range magnetic correlations and not described by the Landau free energy, is commonly observed in frustrated magnets.\cite{Dalmas03} 

\begin{figure}
\centering
\includegraphics[width=0.4\textwidth]{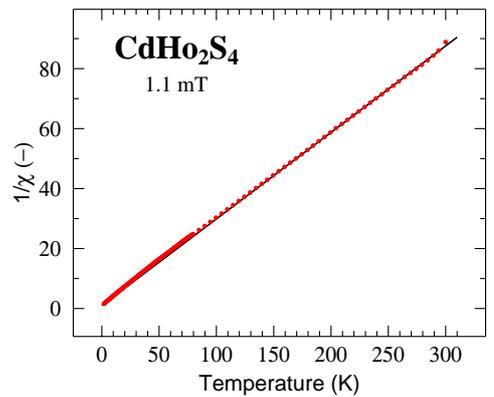}\\
\caption{(Color online) 
Inverse of the magnetic susceptibility versus temperature. The data have been measured in a magnetic field of 1.1~mT. The solid line results from a fit of the Curie-Weiss law to the data measured above 150~K. The susceptibility is dimensionless since we use SI units. 
}
\label{inv_chi}
\end{figure}

We now consider the bulk magnetic measurements. Figure~\ref{inv_chi} depicts the inverse of the susceptibility, i.e.\ $1/\chi$, versus temperature in a broad temperature range. The Curie-Weiss law provides a good description of $\chi$ above 150~K with
$\theta_{\rm CW} = - 3.6 \,(5)$~K and the so-called paramagnetic moment $m_{\rm pm} = 10.8 \, (7) \, \mu_{\rm B}$. This latter number compares favorably with the isolated Ho$^{3+}$ value, $m_{\rm eff} = 10.6  \, \mu_{\rm B}$. From measurements with $B_{\rm ext} = 0.1$~T, Lau {\it et al.} have reported a similar value for $m_{\rm pm}$, but a larger $\theta_{\rm CW}$ absolute value ($\theta_{\rm CW} = - 7.6 \, (2)$~K).\cite{Lau05} Because  $ \theta_{\rm CW}$ is negative, the dominant exchange interactions are antiferromagnetic. A slight deviation from the Curie-Weiss law is observed below 150~K. 

\begin{figure}
\centering
\includegraphics[width=0.4\textwidth]{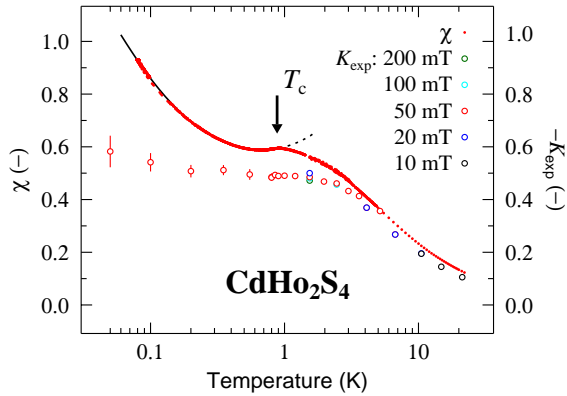}\\
\caption{(Color online) 
Magnetic measurements for a CdHo$_2$S$_4$ powder sample. The data for the magnetic susceptibility $\chi$ versus temperature were recorded with external fields $B_{\rm ext}$ ranging from 0.5 to 50~mT and were found to match one another. The field was applied in the plane of the sample pellet so that the demagnetization field is small. The solid line at low temperature together with the dotted line extension above $T_{\rm c}$ represents a fit of the function $C_{\rm fs}/(T-\theta_{\rm CW, fs}) + a + b\, T$ to the data recorded below $T_{\rm c}$. The former term represents the contribution of weakly interacting Ho$^{3+}$ spins, while the remaining terms describe the majority spin state; see main text. 
The quantity $K_{\rm exp}$ defined in the main text and measured for different $B_{\rm ext}$ as indicated in the figure, is shown for comparison.
}
\label{magnetic_results}
\end{figure}

In Fig.~\ref{magnetic_results}, 
the variation of $\chi$ at low temperature is displayed: it exhibits a weak maximum at $T_{\rm c} \simeq 0.88$~K. This maximum is somewhat stronger in ac\ susceptibility data (not shown) recorded in the 5-100 Hz frequency range. Together with the aforementioned $C_{\rm p}$ peak  these results point to a magnetic transition at $T_{\rm c}$. An uprise in $\chi(T)$ is detected well below $T_{\rm c}$. An origin for it could be the presence of residual free spins in our sample. Modelizing this upturn as explained in the caption of Fig.~\ref{magnetic_results},
a very good fit is obtained for a 1.7\% fraction of Ho$^{3+}$ ions being in a paramagnetic state with a moment $m_{\rm eff}$; see full line in Fig.~\ref{magnetic_results}.
The Curie-Weiss temperature associated with this fraction of the spins is negligible: $\theta_{\rm CW, fs}$ = $-37\,(1)$~mK. 

\begin{figure}
\centering
\includegraphics[width=0.4\textwidth]{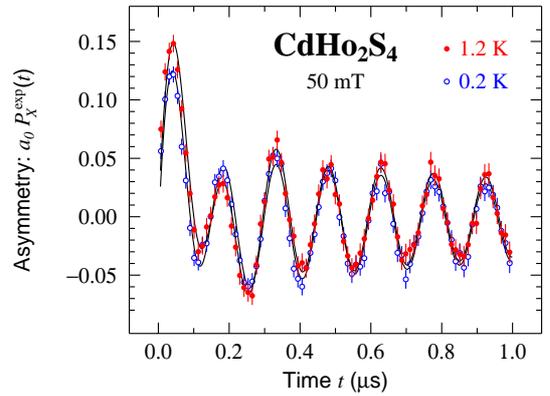}\\
\caption{(Color online) $\mu$SR spectra recorded in a transverse of 50~mT at 0.2 and 1.2~K. The full lines are the results of fits to the data. The model consists of a sum of two exponentially damped cosine functions, the former and latter accounting for muons stopped in the sample and its surroundigs, respectively. The amplitudes extracted from the fits are temperature independent for the two components. The precession frequency $\nu_\mu$ of the former provides a measure of the mean field $\langle B_\mu\rangle$ = $2\pi\nu_\mu/\gamma_\mu$ at the muon site in the sample; see main text for details. $\gamma_\mu =851.615 \, {\rm Mrad} \, {\rm s}^{-1} \, {\rm T}^{-1}$ is the muon gyromagnetic ratio.
}
\label{muon_tf_spectra}
\end{figure}

In order to check the hypothesis that the $\chi$ upturn arises from a small fraction of the Ho$^{3+}$ spins, we have used the $\mu$SR technique in the transverse-field geometry. Here, a field ${\bf B}_{\rm ext}$  transverse to the initial polarization of the muon beam is applied to the sample. The muon Larmor precession is then monitored (Fig.~\ref{muon_tf_spectra}), 
yielding the mean field magnitude $\langle B_\mu\rangle$ at the muon site which {\em uniformly} probes the sample volume. The quantity of interest here is $K_{\rm exp} = (\langle B_\mu \rangle - B_{\rm ext})/B_{\rm ext}$. Once corrected for the demagnetization and Lorentz fields,\footnote{According to Eq.~5.66 of Ref.~\onlinecite{Yaouanc11} the local susceptibility $K_\mu$ is related to $K_{\rm exp}$ through the relation $K_\mu$ = $K_{\rm exp} - (1/3 - N^Z)\chi$ which can be applied since the thin pellet shaped sample can be approximated to a strongly oblate ellipsoid of revolution. The relevant demagnetization coefficient $N^Z$ being definitively larger than 1/3, the demagnetization and Lorentz field corrections for $-K_{\rm exp}$ will result in a downward shift in Fig.~\ref{magnetic_results}, proportional to $\chi$, which rules out any low temperature uprise.} this quantity is proportional to a microscopic susceptibility which can be compared to the macroscopic susceptibility discussed above. The proportionality coefficient depends on the muon-system coupling. In Fig.~\ref{magnetic_results}, $-K_{\rm exp}(T)$ is found to track $\chi(T)$ for $T > 3$~K. 
However for $T\lesssim T_{\rm c}$, $-K_{\rm exp}$ only weakly depends on the temperature. This observation confirms that the uprise of $\chi(T)$ is not representative of the vast majority of the Ho$^{3+}$ spins in CdHo$_2$S$_4$. 

\begin{figure}
\includegraphics[width=0.375\textwidth]{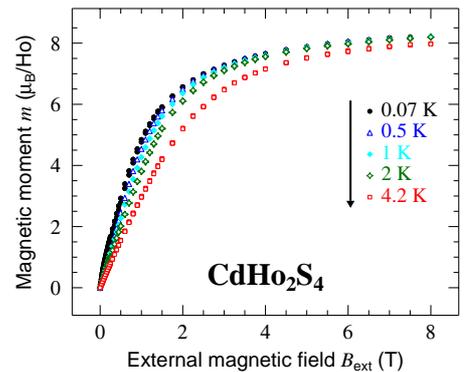}
\caption{(Color online)
Magnetic moment per holmium ion of a CdHo$_2$S$_4$ powder versus field measured for different temperatures as indicated in the figure.
}
\label{magnetization}
\end{figure}

\begin{figure}
\includegraphics[width=0.375\textwidth]{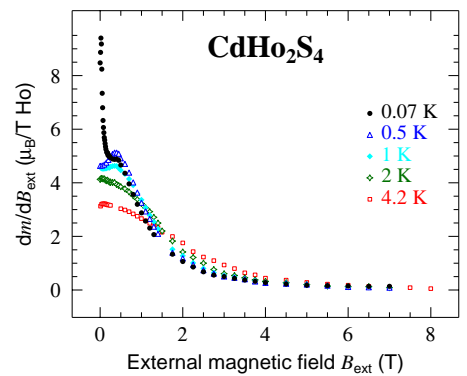}
\caption{(Color online)
External field derivative of the Ho$^{3+}$ magnetic moment of a CdHo$_2$S$_4$ powder for different temperatures as indicated in the figure.
}
\label{magnetization_derivative}
\end{figure}

The measurements of the magnetic moment per holmium ion as a function of field at 4.2~K and down to 0.07~K are presented in Fig.~\ref{magnetization}. The moment tends to saturation with $m_{\rm sat} \simeq 8.2 \, \mu_{\rm B}$/Ho$^{3+}$ under 8~T, fairly consistent with $m_{\rm sat} \simeq 7.5 \, \mu_{\rm B}$/Ho$^{3+}$ at 5~T previously reported.\cite{Lau05}
The value for $m_{\rm sat}$ is much larger in CdHo$_2$S$_4$ than in Ho$_2$Ti$_2$O$_7$,\cite{Petrenko03} suggesting a different type of anisotropy in the two compounds.
With the presently available experimental data the electronic configuration of the Ho$^{3+}$ spins, in particular their crystal electric field energy levels and wave functions cannot be discussed further. In Fig.~\ref{magnetization_derivative} we display the derivative of the magnetic moment value with repect to the external field. 
At 4.2 and 2~K it monotonically decays as the external field is increased. At lower temperatures the derivative passes through a maximum at a field of approximately half a tesla. This maximum is the signature of a metamagnetic-like behavior, as usually observed in antiferromagnets. The sharp decrease of the derivative for fields up to $\simeq 0.2$~T at 0.07~K is more surprising. While we have no definitive explanation for it at the moment, it could be associated with the fraction of free spins detected in the susceptibility measurements.

\begin{figure}
\begin{center}
\includegraphics[width=0.375\textwidth]{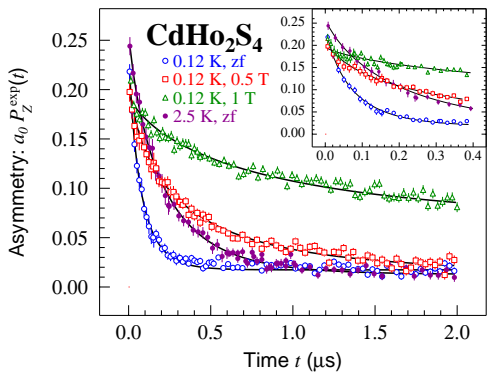}
\end{center}
\caption{(Color online)
Four $\mu$SR spectra recorded for a CdHo$_2$S$_4$ powder sample,
two zero-field spectra taken on both sides of the magnetic phase
transition temperature $T_{\rm c}$, and two longitudinal field spectra recorded at $T=0.12$~K i.e.\ $ T \ll T_{\rm c}$. The early time details are shown in the insert. The solid lines result from fits as explained in the main text.
}
\label{muon_spectra}
\end{figure}

We now discuss the zero-field $\mu$SR spectra. Examples of spectra recorded on both sides of $T_{\rm c}$ are displayed in Fig.~\ref{muon_spectra}.
Contrary to expectation for an ordered magnet, no spontaneous oscillation is detected 
below $T_c$. There is also no missing asymmetry which would result from an unresolved
oscillation. We simply find an exponential-like relaxation on each side of $T_c$.
This is a signature of a strong dynamical spin component below $T_{\rm c}$.

\begin{figure}
\includegraphics[width=0.375\textwidth]{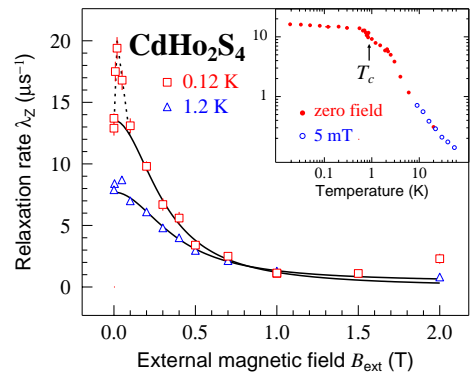}
\caption{(Color online)
Spin-lattice relaxation rate $\lambda_Z$ versus longitudinal field intensity
$B_{\rm ext}$ and temperature $T$ for a CdHo$_2$S$_4$ powder.
In the main frame is displayed  $\lambda_Z(B_{\rm ext})$ for two temperatures below and above $T_{\rm c}$.
The $\lambda_Z (T=0.12~{\rm K})$ maximum occurs at 20~mT.
The solid lines are explained in the main text.
The dashed line at small $B_{\rm ext}$ and $T= 0.12$~K is a guide to the eyes. 
In the insert is displayed $\lambda_Z (T)$
measured from 0.019 to 55~K under zero field or $B_{\rm ext} = 5$~mT. 
The temperature $T_{\rm c}$ at which the compound exhibits a magnetic phase 
transition is specified by an arrow. 
}
\label{muon_field_temperature}
\end{figure}

The function $a_0 P_Z^{\rm exp}(t)= a_{\rm s} P_Z(t) + a_{\rm bg}$, where the second time-independent component accounts for the muons missing the sample, has been fitted to the spectra.
A good description of  $P_Z(t)$ in zero field is obtained with a stretched exponential
relaxation, i.e.\ $P_Z(t) = \exp[-(\lambda_Z t)^\beta]$. The exponent $\beta$
has been found constant with $\beta = 0.8$ up to 0.6~K, and then it increases steadily with temperature, reaching  $\beta =1 $ above 10~K. As seen in the insert of Fig.~\ref{muon_field_temperature}, to the critical temperature corresponds a faint anomaly in $\lambda_Z(T)$. The rate $\lambda_Z$ is finite and becomes almost temperature independent
below $T_{\rm c}$, a signature of the so-called persistent spin dynamics.

The results for CdHo$_2$S$_4$ are consistent with
the previous observation for the ordered spin ice 
Tb$_2$Sn$_2$O$_7$.\cite{Dalmas06} While a spontaneous oscillation
is also absent in the order-by-disorder antiferromagnet Er$_2$Ti$_2$O$_7$ (Refs.~\onlinecite{Lago05,Dalmas12a}) and the 
splayed ferromagnet Yb$_2$Sn$_2$O$_7$,\cite{Yaouanc13} their relaxation below $T_c$ is not 
exponential-like. For Tb$_2$Sn$_2$O$_7$ and CdHo$_2$S$_4$ we are in fact in the fast fluctuation regime for which $\gamma_\mu \Delta_{\rm rms} \, \tau_{\rm c} \ll 1$, where 
$\Delta_{\rm rms}$ and $\tau_{\rm c}$ are the standard deviation of the field distribution at the muon site and the correlation time of the field-correlation function, respectively. {\em This key feature will enable us to investigate the relaxation in terms of spin-correlation functions}. 
Before leaving this qualitative discussion, we note that an inflection point in 
$\lambda_Z(T)$ is present around 20~K. It may correspond to a crossover from
a crystal-electric-field excitation dominated regime\cite{Dalmas03,Yaouanc11} to a strongly 
correlated low-temperature paramagnetic regime. We also note that Tb$_2$Sn$_2$O$_7$ has been discussed in terms of a partially ordered magnet owing to the coexistence of static and dynamical magnetic modes in its ordered phase.\cite{Rule09b} Such a situation may apply for CdHo$_2$S$_4$.

Figure~\ref{muon_spectra} also displays two longitudinal-field spectra recorded at 0.12~K. 
They are reasonably represented by a stretched exponential function. This means that the system is characterized by a distribution of relaxation rates.\cite{Lindsey80,Berderan05,Johnston06,Yaouanc11}
In Fig.~\ref{muon_field_temperature} we present $\lambda_Z(B_{\rm ext})$.
While at 1.2~K it drops monotonically as $B_{\rm ext}$ is increased,
it exhibits a maximum at $\approx 20$~mT for $T= 0.12$~K
before decreasing at larger $B_{\rm ext}$. We also note a slight upturn in $\lambda_Z$ above 1.5~T which could be associated with a crystal-electric-field effect.
A low-field $\lambda_Z$ maximum has already been reported.\cite{Dalmas06,Zorko08,Yaouanc11a,Baker12} An avoided level-crossing resonance might be at play.\cite{Abragam84} A quantitative analysis should provide further information. Neglecting this maximum, we find $\lambda_Z(B_{\rm ext})$ to be well described by a conventional Lorentzian behavior (full lines in Fig.~\ref{muon_field_temperature}): $\lambda_Z(B_{\rm ext}) = 
2 \gamma^2_\mu \Delta_{\rm rms}^2 \tau_{\rm c}/
(1 + \gamma^2_\mu B^2_{\rm ext} \tau^2_{\rm c})$.\cite{Redfield57,Yaouanc11}
At 0.12~K a fit to the data gives $\tau_{\rm c}=3.8\,(3)$~ns and $\Delta_{\rm rms}=49\,(4)$~mT, and at 1.2~K, the parameters are $\tau_{\rm c}=3.08\,(8)$~ns and $\Delta_{\rm rms}=40\,(1)$~mT. 
An additional small constant $\lambda_{Z,0} = 0.4 \, \mu{\rm s}^{-1}$ needs to be added to the Lorentzian at 
1.2 K. Surprisingly, the two parameters of the Lorentzian function have approximately the same values at 0.12 
and 1.2~K. This suggests the same type of excitations to be involved in the relaxation of the muon spin in the 
ordered and paramagnetic states. We also note that the paramagnetic fluctuation time scale of a few nanoseconds is anomalously long: from the energy scale given by the value of $|\Theta_{\rm CW}|$ one would expect a fluctuation time at least two orders of magnitude shorter. 

Before discussing our experimental result in terms of intrinsic properties of the magnetic fluctuation modes that have been uncovered, we note that an alternative explanation has been proposed for the finite and temperature independent relaxation rate measured at low temperature.\cite{Quemerais12} The model put forward by Qu\'emerais and coworkers is based on the coherent diffusion of polaronic muons rather than magnetic fluctuations. However this interpretation of the Dy$_2$Ti$_2$O$_7$ data\cite{Lago07,Dunsiger11} leads to a muon hopping rate nearly three orders of magnitude larger than that measured on the isostructural non-magnetic material Y$_2$Ti$_2$O$_7$.\cite{Rodriguez13} Moreover spin dynamics has been detected in Dy$_2$Ti$_2$O$_7$ down to 0.1~K.\cite{Gardner11} Although no data is available concerning muon diffusion in CdHo$_2$S$_4$, here we will not consider this possibility. Indeed, the similarity of the data in this material, in the pyrochlore systems listed at the beginning of this text, and in Cu$_2$Cl(OH)$_3$ strongly suggests that an explanation generic to three dimensional networks of corner sharing tetrahedral spins must pertain.

To explain the low-temperature finite and approximately temperature independent zero-field $\lambda_Z$ value, a Raman relaxation process involving two magnetic excitations has been put forward.\cite{Yaouanc05a} Generalizing this picture, we write
\begin{eqnarray}
\lambda_Z & = & 
{\mathcal C} \int_\Delta^\infty f[\epsilon/(k_{\rm B} T)] g^2_{\rm m} (\epsilon) {\rm d} \epsilon.
\label{lambdaZ}
\end{eqnarray}
Here ${\mathcal C}$ is a temperature independent constant and $f(x)  =  n(x) [ 1 \pm n(x) ]$, with $n(x)$ the Bose-Einstein or Fermi-Dirac distribution function and the $+$ or $-$ signs are for bosonic or fermionic excitations, respectively. We have introduced the magnetic density of states responsible for the relaxation $g_{\rm m}(\epsilon)$ and an energy gap $\Delta$. To get $\lambda_Z$ temperature independent, we need $g_{\rm m}(\epsilon) = b_\mu \epsilon^{-1/2}$ and $(\Delta -E_{\rm F})$ or $(\Delta - \mu)$ proportional to temperature i.e.\ equal to $a_\mu k_{\rm B} T$, where $a_\mu$ and $b_\mu$ are finite constants. We have denoted $E_{\rm F}$ the Fermi energy and $\mu$ the chemical potential (needed if the boson number is not fixed). The inverse square root form for $g_m(\epsilon)$ needs to be verified only at low energy.\cite{Yaouanc05a}

Expressing $g_{\rm m}(\epsilon)$ in terms of
the spin correlation function $\langle {\bf J}_{\bf q}(t) {\bf J}_{-{\bf q}}(0) \rangle$, we obtain
\begin{eqnarray}
g_{\rm m} (\epsilon) =  \sum _{\bf q}
\int_{-\infty}^\infty 
{\langle {\bf J}_{\bf q}(t) \cdot {\bf J}_{-{\bf q}}(0) \rangle \over 
\langle {\bf J}_{\bf q}(0) \cdot {\bf J}_{-{\bf q}}(0) \rangle }
\exp\left({i \epsilon t \over\hbar}\right) {{\rm d} t  \over 2 \pi \hbar}. 
\label{density}
\end{eqnarray}
The sum is over the first Brillouin zone vectors. 
We recall  $ \langle {\bf J}_{\bf q}(t) \cdot {\bf J}_{-{\bf q}}(0) \rangle = 
\sum_i \exp(-i {\bf q} \cdot {\bf i}) \langle {\bf J}_0(t) \cdot {\bf J}_i(0) \rangle $, where ${\bf J}_i$ and ${\bf J}_0$ are the spins at the lattice point $i$ and at the origin of the lattice, respectively. Since muons probe the very low energy spin excitations it is justified to consider the correlation function at long times. In this limit it is governed by a diffusion equation for a Heisenberg Hamiltonian system,\cite{Bloembergen49,vanHove54,deJongh90} i.e.\ $\langle {\bf J}_0(t) \cdot {\bf J}_i(0) \rangle \propto 1/(D |t|)^{d/2}$
where $d$ is the dimensionality of the spin system and $D$ a diffusion coefficient.
Since the Fourier transform of $\sqrt{1/|t|}$ is $\sqrt{2\pi/|\omega|}$ = $\sqrt{2\pi\hbar/|\varepsilon|}$, unidimensional spin structures, i.e.\ $d =1$, are inferred to explain the  
low-temperature $\lambda_Z$ plateau found for geometrically frustrated magnetic materials,
at least when the relaxation is exponential-like. Although the algebraic decay was originally
derived from a phenomenological high-temperature theory for a Heisenberg system,\cite{Bloembergen49} 
a microscopic analysis leads to the same form for the correlations at low temperature in one-dimensional quantum spin systems with anisotropic exchange.\cite{Sirker11}

We {\em tentatively} associate the low energy unidimensional excitations inferred from the temperature independent spin-lattice relaxation rate observed in numerous frustrated magnets to loop spin structures that have been considered theoretically (see e.g. Refs.~\onlinecite{Villain79,Hermele04, Melko04} for three dimensional systems) and suggested from neutron scattering experiments. An hexamer pattern was reported in the paramagnetic phase of the spinel chromite $A$Cr$_2$O$_4$ with $A$ = Zn,\cite{Lee02}
then for $A$ = Cd,\cite{Chung05} and later on for $A$ = Mg in the paramagnetic and ordered magnetic 
states.\cite{Tomiyasu08,Tomiyasu13} Results for the itinerant system 
Y$_{0.97}$Sc$_{0.03}$Mn$_2$ display a broadly similar feature.\cite{Ballou96}
Hexamer correlations seem also present in the spin-ice system Dy$_2$Ti$_2$O$_7$.\cite{Yavors08} An illustration for an hexamer structure is presented in Fig.~\ref{Structure_specific_heat_data}a, as well as another putative structure.

The similarity of $\lambda_Z(B_{\rm ext})$ measured for CdHo$_2$S$_4$ on both sites of the transition shows unidimensional excitations to be present in both phases. The unusual long time scale observed for the spin dynamics is then consistent with the relatively large number of spins involved in the process. A possible origin for the drop of $\lambda_Z$ above $\approx 1$~K might be 
excitation interactions which would break the spin-correlation power-law decay.

The ubiquituous low-temperature excitations that we infer suggest a generic mechanism. Emergent monopoles are the low-temperature magnetic excitations of spin-ice systems, i.e.\ of ferromagnetically interacting spins on a lattice of corner-sharing tetrahedra.\cite{Ryzhkin05,Castelnovo08} An all-in all-out spin arrangement for the same lattice can be viewed as a lattice of magnetic octupoles.\cite{Arima13} Although still not proven, the same picture implying multipoles may apply to other antiferromagnetic structures. These poles interact through an effective, possibly screened, Coulomb interaction. We suggest to investigate whether this interaction could explain the excitations we have uncovered.

In summary, the thiospinel CdHo$_2$S$_4$ in which the rare earth spins form a lattice of corner sharing regular tetrahedra undergoes a magnetic transition at $T_{\rm c} \simeq$ 0.87~K. A signature of anomalously slow fluctuation modes is found in the paramagnetic state. Similar modes are present in the ordered phase. Spin dynamics is observed down to at least $T_{\rm c}/50$ through a finite and roughly temperature independent muon spin-lattice relaxation rate $\lambda_Z$. This is shown to be the signature of emergent unidimensional spin excitations. Since the $\lambda_Z$ plateau has been found in virtually all the frustrated compounds including the pyrochlore, kagome and triangular systems 
\footnote{See, e.g. Refs.~\onlinecite{Zheng05,Yaouanc05a} for the lattice of corner sharing tetrahedra, Ref.~\onlinecite{Mendels07} for the kagome lattice and Ref.~ \onlinecite{Dalmas12} for the triangular lattice}, we conjecture 
the unidimensional spin excitations to be a generic feature of geometrically frustrated magnets, should the system order or not. 

We thank C. Paulsen for the use of his SQUID dilution magnetometer. PDR gratefully acknowledges partial support from Prof.\ H.\ Keller from the University of Zurich. This research project has been partially supported by the European Commission under the 7th Framework Programme through the `Research Infrastructures' action of the `Capacities' Programme, Contract No: CP-CSA\_INFRA-2008-1.1.1 Number 226507-NMI3. Part of this work was performed at the Swiss Muon Source, Paul Scherrer Institute, Villigen, Switzerland.

\bibliography{reference}

\begin{thebibliography}{57}%
\makeatletter
\providecommand \@ifxundefined [1]{%
 \@ifx{#1\undefined}
}%
\providecommand \@ifnum [1]{%
 \ifnum #1\expandafter \@firstoftwo
 \else \expandafter \@secondoftwo
 \fi
}%
\providecommand \@ifx [1]{%
 \ifx #1\expandafter \@firstoftwo
 \else \expandafter \@secondoftwo
 \fi
}%
\providecommand \natexlab [1]{#1}%
\providecommand \enquote  [1]{``#1''}%
\providecommand \bibnamefont  [1]{#1}%
\providecommand \bibfnamefont [1]{#1}%
\providecommand \citenamefont [1]{#1}%
\providecommand \href@noop [0]{\@secondoftwo}%
\providecommand \href [0]{\begingroup \@sanitize@url \@href}%
\providecommand \@href[1]{\@@startlink{#1}\@@href}%
\providecommand \@@href[1]{\endgroup#1\@@endlink}%
\providecommand \@sanitize@url [0]{\catcode `\\12\catcode `\$12\catcode
  `\&12\catcode `\#12\catcode `\^12\catcode `\_12\catcode `\%12\relax}%
\providecommand \@@startlink[1]{}%
\providecommand \@@endlink[0]{}%
\providecommand \url  [0]{\begingroup\@sanitize@url \@url }%
\providecommand \@url [1]{\endgroup\@href {#1}{\urlprefix }}%
\providecommand \urlprefix  [0]{URL }%
\providecommand \Eprint [0]{\href }%
\providecommand \doibase [0]{http://dx.doi.org/}%
\providecommand \selectlanguage [0]{\@gobble}%
\providecommand \bibinfo  [0]{\@secondoftwo}%
\providecommand \bibfield  [0]{\@secondoftwo}%
\providecommand \translation [1]{[#1]}%
\providecommand \BibitemOpen [0]{}%
\providecommand \bibitemStop [0]{}%
\providecommand \bibitemNoStop [0]{.\EOS\space}%
\providecommand \EOS [0]{\spacefactor3000\relax}%
\providecommand \BibitemShut  [1]{\csname bibitem#1\endcsname}%
\let\auto@bib@innerbib\@empty
\bibitem [{\citenamefont {Gardner}\ \emph {et~al.}(2010)\citenamefont
  {Gardner}, \citenamefont {Gingras},\ and\ \citenamefont
  {Greedan}}]{Gardner10}%
  \BibitemOpen
  \bibfield  {author} {\bibinfo {author} {\bibfnamefont {J.~S.}\ \bibnamefont
  {Gardner}}, \bibinfo {author} {\bibfnamefont {M.~J.~P.}\ \bibnamefont
  {Gingras}}, \ and\ \bibinfo {author} {\bibfnamefont {J.~E.}\ \bibnamefont
  {Greedan}},\ }\href {\doibase 10.1103/RevModPhys.82.53} {\bibfield  {journal}
  {\bibinfo  {journal} {Rev. Mod. Phys.}\ }\textbf {\bibinfo {volume} {82}},\
  \bibinfo {pages} {53} (\bibinfo {year} {2010})}\BibitemShut {NoStop}%
\bibitem [{\citenamefont {Harris}\ \emph {et~al.}(1997)\citenamefont {Harris},
  \citenamefont {Bramwell}, \citenamefont {McMorrow}, \citenamefont {Zeiske},\
  and\ \citenamefont {Godfrey}}]{Harris97}%
  \BibitemOpen
  \bibfield  {author} {\bibinfo {author} {\bibfnamefont {M.~J.}\ \bibnamefont
  {Harris}}, \bibinfo {author} {\bibfnamefont {S.~T.}\ \bibnamefont
  {Bramwell}}, \bibinfo {author} {\bibfnamefont {D.~F.}\ \bibnamefont
  {McMorrow}}, \bibinfo {author} {\bibfnamefont {T.}~\bibnamefont {Zeiske}}, \
  and\ \bibinfo {author} {\bibfnamefont {K.~W.}\ \bibnamefont {Godfrey}},\
  }\href {\doibase 10.1103/PhysRevLett.79.2554} {\bibfield  {journal} {\bibinfo
   {journal} {Phys. Rev. Lett.}\ }\textbf {\bibinfo {volume} {79}},\ \bibinfo
  {pages} {2554} (\bibinfo {year} {1997})}\BibitemShut {NoStop}%
\bibitem [{\citenamefont {Hodges}\ \emph {et~al.}(2002)\citenamefont {Hodges},
  \citenamefont {Bonville}, \citenamefont {Forget}, \citenamefont {Yaouanc},
  \citenamefont {{Dalmas de R\'eotier}}, \citenamefont {Andr\'e}, \citenamefont
  {Rams}, \citenamefont {Kr\'olas}, \citenamefont {Ritter}, \citenamefont
  {Gubbens}, \citenamefont {Kaiser}, \citenamefont {King},\ and\ \citenamefont
  {Baines}}]{Hodges02}%
  \BibitemOpen
  \bibfield  {author} {\bibinfo {author} {\bibfnamefont {J.~A.}\ \bibnamefont
  {Hodges}}, \bibinfo {author} {\bibfnamefont {P.}~\bibnamefont {Bonville}},
  \bibinfo {author} {\bibfnamefont {A.}~\bibnamefont {Forget}}, \bibinfo
  {author} {\bibfnamefont {A.}~\bibnamefont {Yaouanc}}, \bibinfo {author}
  {\bibfnamefont {P.}~\bibnamefont {{Dalmas de R\'eotier}}}, \bibinfo {author}
  {\bibfnamefont {G.}~\bibnamefont {Andr\'e}}, \bibinfo {author} {\bibfnamefont
  {M.}~\bibnamefont {Rams}}, \bibinfo {author} {\bibfnamefont {K.}~\bibnamefont
  {Kr\'olas}}, \bibinfo {author} {\bibfnamefont {C.}~\bibnamefont {Ritter}},
  \bibinfo {author} {\bibfnamefont {P.~C.~M.}\ \bibnamefont {Gubbens}},
  \bibinfo {author} {\bibfnamefont {C.~T.}\ \bibnamefont {Kaiser}}, \bibinfo
  {author} {\bibfnamefont {P.~J.~C.}\ \bibnamefont {King}}, \ and\ \bibinfo
  {author} {\bibfnamefont {C.}~\bibnamefont {Baines}},\ }\href {\doibase
  10.1103/PhysRevLett.88.077204} {\bibfield  {journal} {\bibinfo  {journal}
  {Phys. Rev. Lett.}\ }\textbf {\bibinfo {volume} {88}},\ \bibinfo {pages}
  {077204} (\bibinfo {year} {2002})}\BibitemShut {NoStop}%
\bibitem [{\citenamefont {Yaouanc}\ \emph
  {et~al.}(2011{\natexlab{a}})\citenamefont {Yaouanc}, \citenamefont {{Dalmas
  de R\'eotier}}, \citenamefont {Marin},\ and\ \citenamefont
  {Glazkov}}]{Yaouanc11c}%
  \BibitemOpen
  \bibfield  {author} {\bibinfo {author} {\bibfnamefont {A.}~\bibnamefont
  {Yaouanc}}, \bibinfo {author} {\bibfnamefont {P.}~\bibnamefont {{Dalmas de
  R\'eotier}}}, \bibinfo {author} {\bibfnamefont {C.}~\bibnamefont {Marin}}, \
  and\ \bibinfo {author} {\bibfnamefont {V.}~\bibnamefont {Glazkov}},\ }\href
  {\doibase 10.1103/PhysRevB.84.172408} {\bibfield  {journal} {\bibinfo
  {journal} {Phys. Rev. B}\ }\textbf {\bibinfo {volume} {84}},\ \bibinfo
  {pages} {172408} (\bibinfo {year} {2011}{\natexlab{a}})}\BibitemShut
  {NoStop}%
\bibitem [{\citenamefont {Ross}\ \emph {et~al.}(2011)\citenamefont {Ross},
  \citenamefont {Yaraskavitch}, \citenamefont {Laver}, \citenamefont {Gardner},
  \citenamefont {Quilliam}, \citenamefont {Meng}, \citenamefont {Kycia},
  \citenamefont {Singh}, \citenamefont {Proffen}, \citenamefont {Dabkowska},\
  and\ \citenamefont {Gaulin}}]{Ross11a}%
  \BibitemOpen
  \bibfield  {author} {\bibinfo {author} {\bibfnamefont {K.~A.}\ \bibnamefont
  {Ross}}, \bibinfo {author} {\bibfnamefont {L.~R.}\ \bibnamefont
  {Yaraskavitch}}, \bibinfo {author} {\bibfnamefont {M.}~\bibnamefont {Laver}},
  \bibinfo {author} {\bibfnamefont {J.~S.}\ \bibnamefont {Gardner}}, \bibinfo
  {author} {\bibfnamefont {J.~A.}\ \bibnamefont {Quilliam}}, \bibinfo {author}
  {\bibfnamefont {S.}~\bibnamefont {Meng}}, \bibinfo {author} {\bibfnamefont
  {J.~B.}\ \bibnamefont {Kycia}}, \bibinfo {author} {\bibfnamefont {D.~K.}\
  \bibnamefont {Singh}}, \bibinfo {author} {\bibfnamefont {T.}~\bibnamefont
  {Proffen}}, \bibinfo {author} {\bibfnamefont {H.~A.}\ \bibnamefont
  {Dabkowska}}, \ and\ \bibinfo {author} {\bibfnamefont {B.~D.}\ \bibnamefont
  {Gaulin}},\ }\href {\doibase 10.1103/PhysRevB.84.174442} {\bibfield
  {journal} {\bibinfo  {journal} {Phys. Rev. B}\ }\textbf {\bibinfo {volume}
  {84}},\ \bibinfo {pages} {174442} (\bibinfo {year} {2011})}\BibitemShut
  {NoStop}%
\bibitem [{\citenamefont {Balents}(2010)}]{Balents10}%
  \BibitemOpen
  \bibfield  {author} {\bibinfo {author} {\bibfnamefont {L.}~\bibnamefont
  {Balents}},\ }\href {http://dx.doi.org/10.1038/nature08917} {\bibfield
  {journal} {\bibinfo  {journal} {Nature}\ }\textbf {\bibinfo {volume} {464}},\
  \bibinfo {pages} {199} (\bibinfo {year} {2010})}\BibitemShut {NoStop}%
\bibitem [{\citenamefont {Yaouanc}\ \emph {et~al.}(2013)\citenamefont
  {Yaouanc}, \citenamefont {Dalmas~de R\'eotier}, \citenamefont {Bonville},
  \citenamefont {Hodges}, \citenamefont {Glazkov}, \citenamefont {Keller},
  \citenamefont {Sikolenko}, \citenamefont {Bartkowiak}, \citenamefont {Amato},
  \citenamefont {Baines}, \citenamefont {King}, \citenamefont {Gubbens},\ and\
  \citenamefont {Forget}}]{Yaouanc13}%
  \BibitemOpen
  \bibfield  {author} {\bibinfo {author} {\bibfnamefont {A.}~\bibnamefont
  {Yaouanc}}, \bibinfo {author} {\bibfnamefont {P.}~\bibnamefont {Dalmas~de
  R\'eotier}}, \bibinfo {author} {\bibfnamefont {P.}~\bibnamefont {Bonville}},
  \bibinfo {author} {\bibfnamefont {J.~A.}\ \bibnamefont {Hodges}}, \bibinfo
  {author} {\bibfnamefont {V.}~\bibnamefont {Glazkov}}, \bibinfo {author}
  {\bibfnamefont {L.}~\bibnamefont {Keller}}, \bibinfo {author} {\bibfnamefont
  {V.}~\bibnamefont {Sikolenko}}, \bibinfo {author} {\bibfnamefont
  {M.}~\bibnamefont {Bartkowiak}}, \bibinfo {author} {\bibfnamefont
  {A.}~\bibnamefont {Amato}}, \bibinfo {author} {\bibfnamefont
  {C.}~\bibnamefont {Baines}}, \bibinfo {author} {\bibfnamefont {P.~J.~C.}\
  \bibnamefont {King}}, \bibinfo {author} {\bibfnamefont {P.~C.~M.}\
  \bibnamefont {Gubbens}}, \ and\ \bibinfo {author} {\bibfnamefont
  {A.}~\bibnamefont {Forget}},\ }\href {\doibase
  10.1103/PhysRevLett.110.127207} {\bibfield  {journal} {\bibinfo  {journal}
  {Phys. Rev. Lett.}\ }\textbf {\bibinfo {volume} {110}},\ \bibinfo {pages}
  {127207} (\bibinfo {year} {2013})}\BibitemShut {NoStop}%
\bibitem [{\citenamefont {Savary}\ and\ \citenamefont
  {Balents}(2012)}]{Savary12}%
  \BibitemOpen
  \bibfield  {author} {\bibinfo {author} {\bibfnamefont {L.}~\bibnamefont
  {Savary}}\ and\ \bibinfo {author} {\bibfnamefont {L.}~\bibnamefont
  {Balents}},\ }\href {\doibase 10.1103/PhysRevLett.108.037202} {\bibfield
  {journal} {\bibinfo  {journal} {Phys. Rev. Lett.}\ }\textbf {\bibinfo
  {volume} {108}},\ \bibinfo {pages} {037202} (\bibinfo {year}
  {2012})}\BibitemShut {NoStop}%
\bibitem [{\citenamefont {Zheng}\ \emph {et~al.}(2005)\citenamefont {Zheng},
  \citenamefont {Kubozono}, \citenamefont {Nishiyama}, \citenamefont
  {Higemoto}, \citenamefont {Kawae}, \citenamefont {Koda},\ and\ \citenamefont
  {Xu}}]{Zheng05}%
  \BibitemOpen
  \bibfield  {author} {\bibinfo {author} {\bibfnamefont {X.~G.}\ \bibnamefont
  {Zheng}}, \bibinfo {author} {\bibfnamefont {H.}~\bibnamefont {Kubozono}},
  \bibinfo {author} {\bibfnamefont {K.}~\bibnamefont {Nishiyama}}, \bibinfo
  {author} {\bibfnamefont {W.}~\bibnamefont {Higemoto}}, \bibinfo {author}
  {\bibfnamefont {T.}~\bibnamefont {Kawae}}, \bibinfo {author} {\bibfnamefont
  {A.}~\bibnamefont {Koda}}, \ and\ \bibinfo {author} {\bibfnamefont {C.~N.}\
  \bibnamefont {Xu}},\ }\href {\doibase 10.1103/PhysRevLett.95.057201}
  {\bibfield  {journal} {\bibinfo  {journal} {Phys. Rev. Lett.}\ }\textbf
  {\bibinfo {volume} {95}},\ \bibinfo {pages} {057201} (\bibinfo {year}
  {2005})}\BibitemShut {NoStop}%
\bibitem [{\citenamefont {{Dalmas de R\'eotier}}\ \emph
  {et~al.}(2004)\citenamefont {{Dalmas de R\'eotier}}, \citenamefont
  {Gubbens},\ and\ \citenamefont {Yaouanc}}]{Dalmas04}%
  \BibitemOpen
  \bibfield  {author} {\bibinfo {author} {\bibfnamefont {P.}~\bibnamefont
  {{Dalmas de R\'eotier}}}, \bibinfo {author} {\bibfnamefont {P.~C.~M.}\
  \bibnamefont {Gubbens}}, \ and\ \bibinfo {author} {\bibfnamefont
  {A.}~\bibnamefont {Yaouanc}},\ }\href {\doibase 10.1088/0953-8984/16/40/014}
  {\bibfield  {journal} {\bibinfo  {journal} {J. Phys.: Condens. Matter}\
  }\textbf {\bibinfo {volume} {16}},\ \bibinfo {pages} {S4687} (\bibinfo {year}
  {2004})}\BibitemShut {NoStop}%
\bibitem [{\citenamefont {Chapuis}\ \emph {et~al.}(2009)\citenamefont
  {Chapuis}, \citenamefont {Dalmas~de R\'eotier}, \citenamefont {Marin},
  \citenamefont {Yaouanc}, \citenamefont {Forget}, \citenamefont {Amato},\ and\
  \citenamefont {Baines}}]{Chapuis09b}%
  \BibitemOpen
  \bibfield  {author} {\bibinfo {author} {\bibfnamefont {Y.}~\bibnamefont
  {Chapuis}}, \bibinfo {author} {\bibfnamefont {P.}~\bibnamefont {Dalmas~de
  R\'eotier}}, \bibinfo {author} {\bibfnamefont {C.}~\bibnamefont {Marin}},
  \bibinfo {author} {\bibfnamefont {A.}~\bibnamefont {Yaouanc}}, \bibinfo
  {author} {\bibfnamefont {A.}~\bibnamefont {Forget}}, \bibinfo {author}
  {\bibfnamefont {A.}~\bibnamefont {Amato}}, \ and\ \bibinfo {author}
  {\bibfnamefont {C.}~\bibnamefont {Baines}},\ }\href {\doibase
  10.1016/j.physb.2008.11.195} {\bibfield  {journal} {\bibinfo  {journal}
  {Physica B}\ }\textbf {\bibinfo {volume} {404}},\ \bibinfo {pages} {686}
  (\bibinfo {year} {2009})}\BibitemShut {NoStop}%
\bibitem [{\citenamefont {Yaouanc}\ \emph {et~al.}(2005)\citenamefont
  {Yaouanc}, \citenamefont {{P. Dalmas de R\'eotier}}, \citenamefont {Glazkov},
  \citenamefont {Marin}, \citenamefont {Bonville}, \citenamefont {Hodges},
  \citenamefont {Gubbens}, \citenamefont {Sakarya},\ and\ \citenamefont
  {Baines}}]{Yaouanc05a}%
  \BibitemOpen
  \bibfield  {author} {\bibinfo {author} {\bibfnamefont {A.}~\bibnamefont
  {Yaouanc}}, \bibinfo {author} {\bibnamefont {{P. Dalmas de R\'eotier}}},
  \bibinfo {author} {\bibfnamefont {V.}~\bibnamefont {Glazkov}}, \bibinfo
  {author} {\bibfnamefont {C.}~\bibnamefont {Marin}}, \bibinfo {author}
  {\bibfnamefont {P.}~\bibnamefont {Bonville}}, \bibinfo {author}
  {\bibfnamefont {J.~A.}\ \bibnamefont {Hodges}}, \bibinfo {author}
  {\bibfnamefont {P.~C.~M.}\ \bibnamefont {Gubbens}}, \bibinfo {author}
  {\bibfnamefont {S.}~\bibnamefont {Sakarya}}, \ and\ \bibinfo {author}
  {\bibfnamefont {C.}~\bibnamefont {Baines}},\ }\href {\doibase
  10.1103/PhysRevLett.95.047203} {\bibfield  {journal} {\bibinfo  {journal}
  {Phys. Rev. Lett.}\ }\textbf {\bibinfo {volume} {95}},\ \bibinfo {eid}
  {047203} (\bibinfo {year} {2005})}\BibitemShut {NoStop}%
\bibitem [{\citenamefont {{Dalmas de R\'eotier}}\ \emph
  {et~al.}(2006)\citenamefont {{Dalmas de R\'eotier}}, \citenamefont {Yaouanc},
  \citenamefont {Keller}, \citenamefont {Cervellino}, \citenamefont {Roessli},
  \citenamefont {Baines}, \citenamefont {Forget}, \citenamefont {Vaju},
  \citenamefont {Gubbens}, \citenamefont {Amato},\ and\ \citenamefont
  {King}}]{Dalmas06}%
  \BibitemOpen
  \bibfield  {author} {\bibinfo {author} {\bibfnamefont {P.}~\bibnamefont
  {{Dalmas de R\'eotier}}}, \bibinfo {author} {\bibfnamefont {A.}~\bibnamefont
  {Yaouanc}}, \bibinfo {author} {\bibfnamefont {L.}~\bibnamefont {Keller}},
  \bibinfo {author} {\bibfnamefont {A.}~\bibnamefont {Cervellino}}, \bibinfo
  {author} {\bibfnamefont {B.}~\bibnamefont {Roessli}}, \bibinfo {author}
  {\bibfnamefont {C.}~\bibnamefont {Baines}}, \bibinfo {author} {\bibfnamefont
  {A.}~\bibnamefont {Forget}}, \bibinfo {author} {\bibfnamefont
  {C.}~\bibnamefont {Vaju}}, \bibinfo {author} {\bibfnamefont {P.~C.~M.}\
  \bibnamefont {Gubbens}}, \bibinfo {author} {\bibfnamefont {A.}~\bibnamefont
  {Amato}}, \ and\ \bibinfo {author} {\bibfnamefont {P.~J.~C.}\ \bibnamefont
  {King}},\ }\href {\doibase 10.1103/PhysRevLett.96.127202} {\bibfield
  {journal} {\bibinfo  {journal} {Phys. Rev. Lett.}\ }\textbf {\bibinfo
  {volume} {96}},\ \bibinfo {pages} {127202} (\bibinfo {year}
  {2006})}\BibitemShut {NoStop}%
\bibitem [{\citenamefont {Bert}\ \emph {et~al.}(2006)\citenamefont {Bert},
  \citenamefont {Mendels}, \citenamefont {Olariu}, \citenamefont {Blanchard},
  \citenamefont {Collin}, \citenamefont {Amato}, \citenamefont {Baines},\ and\
  \citenamefont {Hillier}}]{Bert06}%
  \BibitemOpen
  \bibfield  {author} {\bibinfo {author} {\bibfnamefont {F.}~\bibnamefont
  {Bert}}, \bibinfo {author} {\bibfnamefont {P.}~\bibnamefont {Mendels}},
  \bibinfo {author} {\bibfnamefont {A.}~\bibnamefont {Olariu}}, \bibinfo
  {author} {\bibfnamefont {N.}~\bibnamefont {Blanchard}}, \bibinfo {author}
  {\bibfnamefont {G.}~\bibnamefont {Collin}}, \bibinfo {author} {\bibfnamefont
  {A.}~\bibnamefont {Amato}}, \bibinfo {author} {\bibfnamefont
  {C.}~\bibnamefont {Baines}}, \ and\ \bibinfo {author} {\bibfnamefont {A.~D.}\
  \bibnamefont {Hillier}},\ }\href {\doibase 10.1103/PhysRevLett.97.117203}
  {\bibfield  {journal} {\bibinfo  {journal} {Phys. Rev. Lett.}\ }\textbf
  {\bibinfo {volume} {97}},\ \bibinfo {pages} {117203} (\bibinfo {year}
  {2006})}\BibitemShut {NoStop}%
\bibitem [{\citenamefont {Lago}\ \emph {et~al.}(2005)\citenamefont {Lago},
  \citenamefont {Lancaster}, \citenamefont {Blundell}, \citenamefont
  {Bramwell}, \citenamefont {Pratt}, \citenamefont {Shirai},\ and\
  \citenamefont {Baines}}]{Lago05}%
  \BibitemOpen
  \bibfield  {author} {\bibinfo {author} {\bibfnamefont {J.}~\bibnamefont
  {Lago}}, \bibinfo {author} {\bibfnamefont {T.}~\bibnamefont {Lancaster}},
  \bibinfo {author} {\bibfnamefont {S.~J.}\ \bibnamefont {Blundell}}, \bibinfo
  {author} {\bibfnamefont {S.~T.}\ \bibnamefont {Bramwell}}, \bibinfo {author}
  {\bibfnamefont {F.~L.}\ \bibnamefont {Pratt}}, \bibinfo {author}
  {\bibfnamefont {M.}~\bibnamefont {Shirai}}, \ and\ \bibinfo {author}
  {\bibfnamefont {C.}~\bibnamefont {Baines}},\ }\href
  {http://stacks.iop.org/0953-8984/17/i=6/a=015} {\bibfield  {journal}
  {\bibinfo  {journal} {J. Phys.: Condens. Matter}\ }\textbf {\bibinfo {volume}
  {17}},\ \bibinfo {pages} {979} (\bibinfo {year} {2005})}\BibitemShut
  {NoStop}%
\bibitem [{\citenamefont {Dalmas~de R\'eotier}\ \emph
  {et~al.}(2012{\natexlab{a}})\citenamefont {Dalmas~de R\'eotier},
  \citenamefont {Yaouanc}, \citenamefont {Chapuis}, \citenamefont {Curnoe},
  \citenamefont {Grenier}, \citenamefont {Ressouche}, \citenamefont {Marin},
  \citenamefont {Lago}, \citenamefont {Baines},\ and\ \citenamefont
  {Giblin}}]{Dalmas12a}%
  \BibitemOpen
  \bibfield  {author} {\bibinfo {author} {\bibfnamefont {P.}~\bibnamefont
  {Dalmas~de R\'eotier}}, \bibinfo {author} {\bibfnamefont {A.}~\bibnamefont
  {Yaouanc}}, \bibinfo {author} {\bibfnamefont {Y.}~\bibnamefont {Chapuis}},
  \bibinfo {author} {\bibfnamefont {S.~H.}\ \bibnamefont {Curnoe}}, \bibinfo
  {author} {\bibfnamefont {B.}~\bibnamefont {Grenier}}, \bibinfo {author}
  {\bibfnamefont {E.}~\bibnamefont {Ressouche}}, \bibinfo {author}
  {\bibfnamefont {C.}~\bibnamefont {Marin}}, \bibinfo {author} {\bibfnamefont
  {J.}~\bibnamefont {Lago}}, \bibinfo {author} {\bibfnamefont {C.}~\bibnamefont
  {Baines}}, \ and\ \bibinfo {author} {\bibfnamefont {S.~R.}\ \bibnamefont
  {Giblin}},\ }\href {\doibase 10.1103/PhysRevB.86.104424} {\bibfield
  {journal} {\bibinfo  {journal} {Phys. Rev. B}\ }\textbf {\bibinfo {volume}
  {86}},\ \bibinfo {pages} {104424} (\bibinfo {year}
  {2012}{\natexlab{a}})}\BibitemShut {NoStop}%
\bibitem [{\citenamefont {Chapuis}\ \emph {et~al.}(2007)\citenamefont
  {Chapuis}, \citenamefont {Yaouanc}, \citenamefont {{Dalmas de R\'eotier}},
  \citenamefont {Pouget}, \citenamefont {Fouquet}, \citenamefont {Cervellino},\
  and\ \citenamefont {Forget}}]{Chapuis07}%
  \BibitemOpen
  \bibfield  {author} {\bibinfo {author} {\bibfnamefont {Y.}~\bibnamefont
  {Chapuis}}, \bibinfo {author} {\bibfnamefont {A.}~\bibnamefont {Yaouanc}},
  \bibinfo {author} {\bibfnamefont {P.}~\bibnamefont {{Dalmas de R\'eotier}}},
  \bibinfo {author} {\bibfnamefont {S.}~\bibnamefont {Pouget}}, \bibinfo
  {author} {\bibfnamefont {P.}~\bibnamefont {Fouquet}}, \bibinfo {author}
  {\bibfnamefont {A.}~\bibnamefont {Cervellino}}, \ and\ \bibinfo {author}
  {\bibfnamefont {A.}~\bibnamefont {Forget}},\ }\href
  {http://stacks.iop.org/0953-8984/19/i=44/a=446206} {\bibfield  {journal}
  {\bibinfo  {journal} {J. Phys.: Condens. Matter}\ }\textbf {\bibinfo {volume}
  {19}},\ \bibinfo {pages} {446206} (\bibinfo {year} {2007})}\BibitemShut
  {NoStop}%
\bibitem [{\citenamefont {Rule}\ \emph {et~al.}(2009)\citenamefont {Rule},
  \citenamefont {Ehlers}, \citenamefont {Gardner}, \citenamefont {Qiu},
  \citenamefont {Moskvin}, \citenamefont {Kiefer},\ and\ \citenamefont
  {Gerischer}}]{Rule09b}%
  \BibitemOpen
  \bibfield  {author} {\bibinfo {author} {\bibfnamefont {K.~C.}\ \bibnamefont
  {Rule}}, \bibinfo {author} {\bibfnamefont {G.}~\bibnamefont {Ehlers}},
  \bibinfo {author} {\bibfnamefont {J.~S.}\ \bibnamefont {Gardner}}, \bibinfo
  {author} {\bibfnamefont {Y.}~\bibnamefont {Qiu}}, \bibinfo {author}
  {\bibfnamefont {E.}~\bibnamefont {Moskvin}}, \bibinfo {author} {\bibfnamefont
  {K.}~\bibnamefont {Kiefer}}, \ and\ \bibinfo {author} {\bibfnamefont
  {S.}~\bibnamefont {Gerischer}},\ }\href
  {http://stacks.iop.org/0953-8984/21/i=48/a=486005} {\bibfield  {journal}
  {\bibinfo  {journal} {J. Phys.: Condens. Matter}\ }\textbf {\bibinfo {volume}
  {21}},\ \bibinfo {pages} {486005} (\bibinfo {year} {2009})}\BibitemShut
  {NoStop}%
\bibitem [{\citenamefont {Lau}\ \emph {et~al.}(2005)\citenamefont {Lau},
  \citenamefont {Freitas}, \citenamefont {Ueland}, \citenamefont {Schiffer},\
  and\ \citenamefont {Cava}}]{Lau05}%
  \BibitemOpen
  \bibfield  {author} {\bibinfo {author} {\bibfnamefont {G.~C.}\ \bibnamefont
  {Lau}}, \bibinfo {author} {\bibfnamefont {R.~S.}\ \bibnamefont {Freitas}},
  \bibinfo {author} {\bibfnamefont {B.~G.}\ \bibnamefont {Ueland}}, \bibinfo
  {author} {\bibfnamefont {P.}~\bibnamefont {Schiffer}}, \ and\ \bibinfo
  {author} {\bibfnamefont {R.~J.}\ \bibnamefont {Cava}},\ }\href {\doibase
  10.1103/PhysRevB.72.054411} {\bibfield  {journal} {\bibinfo  {journal} {Phys.
  Rev. B}\ }\textbf {\bibinfo {volume} {72}},\ \bibinfo {pages} {054411}
  (\bibinfo {year} {2005})}\BibitemShut {NoStop}%
\bibitem [{\citenamefont {Lago}\ \emph {et~al.}(2010)\citenamefont {Lago},
  \citenamefont {\ifmmode \check{Z}\else
  \v{Z}\fi{}ivkovi\ifmmode~\acute{c}\else \'{c}\fi{}}, \citenamefont {Malkin},
  \citenamefont {Rodriguez~Fernandez}, \citenamefont {Ghigna}, \citenamefont
  {Dalmas~de R\'eotier}, \citenamefont {Yaouanc},\ and\ \citenamefont
  {Rojo}}]{Lago10}%
  \BibitemOpen
  \bibfield  {author} {\bibinfo {author} {\bibfnamefont {J.}~\bibnamefont
  {Lago}}, \bibinfo {author} {\bibfnamefont {I.}~\bibnamefont {\ifmmode
  \check{Z}\else \v{Z}\fi{}ivkovi\ifmmode~\acute{c}\else \'{c}\fi{}}}, \bibinfo
  {author} {\bibfnamefont {B.~Z.}\ \bibnamefont {Malkin}}, \bibinfo {author}
  {\bibfnamefont {J.}~\bibnamefont {Rodriguez~Fernandez}}, \bibinfo {author}
  {\bibfnamefont {P.}~\bibnamefont {Ghigna}}, \bibinfo {author} {\bibfnamefont
  {P.}~\bibnamefont {Dalmas~de R\'eotier}}, \bibinfo {author} {\bibfnamefont
  {A.}~\bibnamefont {Yaouanc}}, \ and\ \bibinfo {author} {\bibfnamefont
  {T.}~\bibnamefont {Rojo}},\ }\href {\doibase 10.1103/PhysRevLett.104.247203}
  {\bibfield  {journal} {\bibinfo  {journal} {Phys. Rev. Lett.}\ }\textbf
  {\bibinfo {volume} {104}},\ \bibinfo {pages} {247203} (\bibinfo {year}
  {2010})}\BibitemShut {NoStop}%
\bibitem [{\citenamefont {Paulsen}(2001)}]{Paulsen01}%
  \BibitemOpen
  \bibfield  {author} {\bibinfo {author} {\bibfnamefont {C.}~\bibnamefont
  {Paulsen}},\ }in\ \href@noop {} {\emph {\bibinfo {booktitle} {Introduction to
  Physical Techniques in Molecular Magnetism: Structural and Macroscopic
  Techniques -- Yesa 1999}}},\ \bibinfo {editor} {edited by\ \bibinfo {editor}
  {\bibfnamefont {F.}~\bibnamefont {Palacio}}, \bibinfo {editor} {\bibfnamefont
  {E.}~\bibnamefont {Ressouche}}, \ and\ \bibinfo {editor} {\bibfnamefont
  {J.}~\bibnamefont {Schweizer}}}\ (\bibinfo  {publisher} {Servicio de
  Publicaciones de la Universidad de Zaragoza},\ \bibinfo {address}
  {Zaragoza},\ \bibinfo {year} {2001})\BibitemShut {NoStop}%
\bibitem [{\citenamefont {Yaouanc}\ and\ \citenamefont {{Dalmas de
  R\'eotier}}(2011)}]{Yaouanc11}%
  \BibitemOpen
  \bibfield  {author} {\bibinfo {author} {\bibfnamefont {A.}~\bibnamefont
  {Yaouanc}}\ and\ \bibinfo {author} {\bibfnamefont {P.}~\bibnamefont {{Dalmas
  de R\'eotier}}},\ }\href@noop {} {\emph {\bibinfo {title} {Muon Spin
  Rotation, Relaxation, and Resonance: Applications to Condensed Matter}}}\
  (\bibinfo  {publisher} {Oxford University Press},\ \bibinfo {address}
  {Oxford},\ \bibinfo {year} {2011})\BibitemShut {NoStop}%
\bibitem [{\citenamefont {{Dalmas de R\'eotier}}\ \emph
  {et~al.}(2003)\citenamefont {{Dalmas de R\'eotier}}, \citenamefont {Yaouanc},
  \citenamefont {Gubbens}, \citenamefont {Kaiser}, \citenamefont {Baines},\
  and\ \citenamefont {King}}]{Dalmas03}%
  \BibitemOpen
  \bibfield  {author} {\bibinfo {author} {\bibfnamefont {P.}~\bibnamefont
  {{Dalmas de R\'eotier}}}, \bibinfo {author} {\bibfnamefont {A.}~\bibnamefont
  {Yaouanc}}, \bibinfo {author} {\bibfnamefont {P.~C.~M.}\ \bibnamefont
  {Gubbens}}, \bibinfo {author} {\bibfnamefont {C.~T.}\ \bibnamefont {Kaiser}},
  \bibinfo {author} {\bibfnamefont {C.}~\bibnamefont {Baines}}, \ and\ \bibinfo
  {author} {\bibfnamefont {P.~J.~C.}\ \bibnamefont {King}},\ }\href {\doibase
  10.1103/PhysRevLett.91.167201} {\bibfield  {journal} {\bibinfo  {journal}
  {Phys. Rev. Lett.}\ }\textbf {\bibinfo {volume} {91}},\ \bibinfo {pages}
  {167201} (\bibinfo {year} {2003})}\BibitemShut {NoStop}%
\bibitem [{Note1()}]{Note1}%
  \BibitemOpen
  \bibinfo {note} {According to Eq.~5.66 of Ref.~\protect \rev@citealpnum
  {Yaouanc11} the local susceptibility $K_\mu $ is related to $K_{\protect \rm
  exp}$ through the relation $K_\mu $ = $K_{\protect \rm exp} - (1/3 - N^Z)\chi
  $ which can be applied since the thin pellet shaped sample can be
  approximated to a strongly oblate ellipsoid of revolution. The relevant
  demagnetization coefficient $N^Z$ being definitively larger than 1/3, the
  demagnetization and Lorentz field corrections for $-K_{\protect \rm exp}$
  will result in a downward shift in Fig.~\ref {magnetic_results}, proportional
  to $\chi $, which rules out any low temperature uprise.}\BibitemShut {Stop}%
\bibitem [{\citenamefont {Petrenko}\ \emph {et~al.}(2003)\citenamefont
  {Petrenko}, \citenamefont {Lees},\ and\ \citenamefont
  {Balakrishnan}}]{Petrenko03}%
  \BibitemOpen
  \bibfield  {author} {\bibinfo {author} {\bibfnamefont {O.~A.}\ \bibnamefont
  {Petrenko}}, \bibinfo {author} {\bibfnamefont {M.~R.}\ \bibnamefont {Lees}},
  \ and\ \bibinfo {author} {\bibfnamefont {G.}~\bibnamefont {Balakrishnan}},\
  }\href {\doibase 10.1103/PhysRevB.68.012406} {\bibfield  {journal} {\bibinfo
  {journal} {Phys. Rev. B}\ }\textbf {\bibinfo {volume} {68}},\ \bibinfo
  {pages} {012406} (\bibinfo {year} {2003})}\BibitemShut {NoStop}%
\bibitem [{\citenamefont {Lindsey}\ and\ \citenamefont
  {Patterson}(1980)}]{Lindsey80}%
  \BibitemOpen
  \bibfield  {author} {\bibinfo {author} {\bibfnamefont {C.~P.}\ \bibnamefont
  {Lindsey}}\ and\ \bibinfo {author} {\bibfnamefont {G.~D.}\ \bibnamefont
  {Patterson}},\ }\href@noop {} {\bibfield  {journal} {\bibinfo  {journal} {J.
  Chem. Phys.}\ }\textbf {\bibinfo {volume} {73}},\ \bibinfo {pages} {3348}
  (\bibinfo {year} {1980})}\BibitemShut {NoStop}%
\bibitem [{\citenamefont {Berderan-Santos}\ \emph {et~al.}(2005)\citenamefont
  {Berderan-Santos}, \citenamefont {Bodunov},\ and\ \citenamefont
  {Valeur}}]{Berderan05}%
  \BibitemOpen
  \bibfield  {author} {\bibinfo {author} {\bibfnamefont {M.~N.}\ \bibnamefont
  {Berderan-Santos}}, \bibinfo {author} {\bibfnamefont {E.~N.}\ \bibnamefont
  {Bodunov}}, \ and\ \bibinfo {author} {\bibfnamefont {B.}~\bibnamefont
  {Valeur}},\ }\href@noop {} {\bibfield  {journal} {\bibinfo  {journal} {Chem.
  Phys.}\ }\textbf {\bibinfo {volume} {315}},\ \bibinfo {pages} {171} (\bibinfo
  {year} {2005})}\BibitemShut {NoStop}%
\bibitem [{\citenamefont {Johnston}(2006)}]{Johnston06}%
  \BibitemOpen
  \bibfield  {author} {\bibinfo {author} {\bibfnamefont {D.~C.}\ \bibnamefont
  {Johnston}},\ }\href@noop {} {\bibfield  {journal} {\bibinfo  {journal}
  {Phys. Rev. B}\ }\textbf {\bibinfo {volume} {74}},\ \bibinfo {pages} {184430}
  (\bibinfo {year} {2006})}\BibitemShut {NoStop}%
\bibitem [{\citenamefont {Zorko}\ \emph {et~al.}(2008)\citenamefont {Zorko},
  \citenamefont {Bert}, \citenamefont {Mendels}, \citenamefont {Bordet},
  \citenamefont {Lejay},\ and\ \citenamefont {Robert}}]{Zorko08}%
  \BibitemOpen
  \bibfield  {author} {\bibinfo {author} {\bibfnamefont {A.}~\bibnamefont
  {Zorko}}, \bibinfo {author} {\bibfnamefont {F.}~\bibnamefont {Bert}},
  \bibinfo {author} {\bibfnamefont {P.}~\bibnamefont {Mendels}}, \bibinfo
  {author} {\bibfnamefont {P.}~\bibnamefont {Bordet}}, \bibinfo {author}
  {\bibfnamefont {P.}~\bibnamefont {Lejay}}, \ and\ \bibinfo {author}
  {\bibfnamefont {J.}~\bibnamefont {Robert}},\ }\href {\doibase
  10.1103/PhysRevLett.100.147201} {\bibfield  {journal} {\bibinfo  {journal}
  {Phys. Rev. Lett.}\ }\textbf {\bibinfo {volume} {100}},\ \bibinfo {pages}
  {147201} (\bibinfo {year} {2008})}\BibitemShut {NoStop}%
\bibitem [{\citenamefont {Yaouanc}\ \emph
  {et~al.}(2011{\natexlab{b}})\citenamefont {Yaouanc}, \citenamefont {{Dalmas
  de R\'eotier}}, \citenamefont {Chapuis}, \citenamefont {Marin}, \citenamefont
  {Vanishri}, \citenamefont {Aoki}, \citenamefont {F{\aa}k}, \citenamefont
  {Regnault}, \citenamefont {Buisson}, \citenamefont {Amato}, \citenamefont
  {Baines},\ and\ \citenamefont {Hillier}}]{Yaouanc11a}%
  \BibitemOpen
  \bibfield  {author} {\bibinfo {author} {\bibfnamefont {A.}~\bibnamefont
  {Yaouanc}}, \bibinfo {author} {\bibfnamefont {P.}~\bibnamefont {{Dalmas de
  R\'eotier}}}, \bibinfo {author} {\bibfnamefont {Y.}~\bibnamefont {Chapuis}},
  \bibinfo {author} {\bibfnamefont {C.}~\bibnamefont {Marin}}, \bibinfo
  {author} {\bibfnamefont {S.}~\bibnamefont {Vanishri}}, \bibinfo {author}
  {\bibfnamefont {D.}~\bibnamefont {Aoki}}, \bibinfo {author} {\bibfnamefont
  {B.}~\bibnamefont {F{\aa}k}}, \bibinfo {author} {\bibfnamefont {L.~P.}\
  \bibnamefont {Regnault}}, \bibinfo {author} {\bibfnamefont {C.}~\bibnamefont
  {Buisson}}, \bibinfo {author} {\bibfnamefont {A.}~\bibnamefont {Amato}},
  \bibinfo {author} {\bibfnamefont {C.}~\bibnamefont {Baines}}, \ and\ \bibinfo
  {author} {\bibfnamefont {A.~D.}\ \bibnamefont {Hillier}},\ }\href {\doibase
  10.1103/PhysRevB.84.184403} {\bibfield  {journal} {\bibinfo  {journal} {Phys.
  Rev. B}\ }\textbf {\bibinfo {volume} {84}},\ \bibinfo {pages} {184403}
  (\bibinfo {year} {2011}{\natexlab{b}})}\BibitemShut {NoStop}%
\bibitem [{\citenamefont {Baker}\ \emph {et~al.}(2012)\citenamefont {Baker},
  \citenamefont {Matthews}, \citenamefont {Giblin}, \citenamefont {Schiffer},
  \citenamefont {Baines},\ and\ \citenamefont {Prabhakaran}}]{Baker12}%
  \BibitemOpen
  \bibfield  {author} {\bibinfo {author} {\bibfnamefont {P.~J.}\ \bibnamefont
  {Baker}}, \bibinfo {author} {\bibfnamefont {M.~J.}\ \bibnamefont {Matthews}},
  \bibinfo {author} {\bibfnamefont {S.~R.}\ \bibnamefont {Giblin}}, \bibinfo
  {author} {\bibfnamefont {P.}~\bibnamefont {Schiffer}}, \bibinfo {author}
  {\bibfnamefont {C.}~\bibnamefont {Baines}}, \ and\ \bibinfo {author}
  {\bibfnamefont {D.}~\bibnamefont {Prabhakaran}},\ }\href {\doibase
  10.1103/PhysRevB.86.094424} {\bibfield  {journal} {\bibinfo  {journal} {Phys.
  Rev. B}\ }\textbf {\bibinfo {volume} {86}},\ \bibinfo {pages} {094424}
  (\bibinfo {year} {2012})}\BibitemShut {NoStop}%
\bibitem [{\citenamefont {Abragam}(1984)}]{Abragam84}%
  \BibitemOpen
  \bibfield  {author} {\bibinfo {author} {\bibfnamefont {A.}~\bibnamefont
  {Abragam}},\ }\href@noop {} {\bibfield  {journal} {\bibinfo  {journal} {C. R.
  Acad. Sci.}\ }\textbf {\bibinfo {volume} {299}},\ \bibinfo {pages} {95}
  (\bibinfo {year} {1984})}\BibitemShut {NoStop}%
\bibitem [{\citenamefont {Redfield}(1957)}]{Redfield57}%
  \BibitemOpen
  \bibfield  {author} {\bibinfo {author} {\bibfnamefont {A.~G.}\ \bibnamefont
  {Redfield}},\ }\href@noop {} {\bibfield  {journal} {\bibinfo  {journal} {IBM
  J. Research and Development}\ }\textbf {\bibinfo {volume} {1}},\ \bibinfo
  {pages} {19} (\bibinfo {year} {1957})}\BibitemShut {NoStop}%
\bibitem [{\citenamefont {Qu\'emerais}\ \emph {et~al.}(2012)\citenamefont
  {Qu\'emerais}, \citenamefont {McClarty},\ and\ \citenamefont
  {Moessner}}]{Quemerais12}%
  \BibitemOpen
  \bibfield  {author} {\bibinfo {author} {\bibfnamefont {P.}~\bibnamefont
  {Qu\'emerais}}, \bibinfo {author} {\bibfnamefont {P.}~\bibnamefont
  {McClarty}}, \ and\ \bibinfo {author} {\bibfnamefont {R.}~\bibnamefont
  {Moessner}},\ }\href {\doibase 10.1103/PhysRevLett.109.127601} {\bibfield
  {journal} {\bibinfo  {journal} {Phys. Rev. Lett.}\ }\textbf {\bibinfo
  {volume} {109}},\ \bibinfo {pages} {127601} (\bibinfo {year}
  {2012})}\BibitemShut {NoStop}%
\bibitem [{\citenamefont {Lago}\ \emph {et~al.}(2007)\citenamefont {Lago},
  \citenamefont {Blundell},\ and\ \citenamefont {Baines}}]{Lago07}%
  \BibitemOpen
  \bibfield  {author} {\bibinfo {author} {\bibfnamefont {J.}~\bibnamefont
  {Lago}}, \bibinfo {author} {\bibfnamefont {S.~J.}\ \bibnamefont {Blundell}},
  \ and\ \bibinfo {author} {\bibfnamefont {C.}~\bibnamefont {Baines}},\ }\href
  {http://iopscience.iop.org/0953-8984/19/32/326210} {\bibfield  {journal}
  {\bibinfo  {journal} {J. Phys.: Condens. Matter}\ }\textbf {\bibinfo {volume}
  {19}},\ \bibinfo {pages} {326210} (\bibinfo {year} {2007})}\BibitemShut
  {NoStop}%
\bibitem [{\citenamefont {Dunsiger}\ \emph {et~al.}(2011)\citenamefont
  {Dunsiger}, \citenamefont {Aczel}, \citenamefont {Arguello}, \citenamefont
  {Dabkowska}, \citenamefont {Dabkowski}, \citenamefont {Du}, \citenamefont
  {Goko}, \citenamefont {Javanparast}, \citenamefont {Lin}, \citenamefont
  {Ning}, \citenamefont {Noad}, \citenamefont {Singh}, \citenamefont
  {Williams}, \citenamefont {Uemura}, \citenamefont {Gingras},\ and\
  \citenamefont {Luke}}]{Dunsiger11}%
  \BibitemOpen
  \bibfield  {author} {\bibinfo {author} {\bibfnamefont {S.~R.}\ \bibnamefont
  {Dunsiger}}, \bibinfo {author} {\bibfnamefont {A.~A.}\ \bibnamefont {Aczel}},
  \bibinfo {author} {\bibfnamefont {C.}~\bibnamefont {Arguello}}, \bibinfo
  {author} {\bibfnamefont {H.}~\bibnamefont {Dabkowska}}, \bibinfo {author}
  {\bibfnamefont {A.}~\bibnamefont {Dabkowski}}, \bibinfo {author}
  {\bibfnamefont {M.-H.}\ \bibnamefont {Du}}, \bibinfo {author} {\bibfnamefont
  {T.}~\bibnamefont {Goko}}, \bibinfo {author} {\bibfnamefont {B.}~\bibnamefont
  {Javanparast}}, \bibinfo {author} {\bibfnamefont {T.}~\bibnamefont {Lin}},
  \bibinfo {author} {\bibfnamefont {F.~L.}\ \bibnamefont {Ning}}, \bibinfo
  {author} {\bibfnamefont {H.~M.~L.}\ \bibnamefont {Noad}}, \bibinfo {author}
  {\bibfnamefont {D.~J.}\ \bibnamefont {Singh}}, \bibinfo {author}
  {\bibfnamefont {T.~J.}\ \bibnamefont {Williams}}, \bibinfo {author}
  {\bibfnamefont {Y.~J.}\ \bibnamefont {Uemura}}, \bibinfo {author}
  {\bibfnamefont {M.~J.~P.}\ \bibnamefont {Gingras}}, \ and\ \bibinfo {author}
  {\bibfnamefont {G.~M.}\ \bibnamefont {Luke}},\ }\href {\doibase
  10.1103/PhysRevLett.107.207207} {\bibfield  {journal} {\bibinfo  {journal}
  {Phys. Rev. Lett.}\ }\textbf {\bibinfo {volume} {107}},\ \bibinfo {pages}
  {207207} (\bibinfo {year} {2011})}\BibitemShut {NoStop}%
\bibitem [{\citenamefont {Rodriguez}\ \emph {et~al.}(2013)\citenamefont
  {Rodriguez}, \citenamefont {Yaouanc}, \citenamefont {Barbara}, \citenamefont
  {Pomjakushina}, \citenamefont {{Qu\'emerais}},\ and\ \citenamefont
  {Salman}}]{Rodriguez13}%
  \BibitemOpen
  \bibfield  {author} {\bibinfo {author} {\bibfnamefont {J.~A.}\ \bibnamefont
  {Rodriguez}}, \bibinfo {author} {\bibfnamefont {A.}~\bibnamefont {Yaouanc}},
  \bibinfo {author} {\bibfnamefont {B.}~\bibnamefont {Barbara}}, \bibinfo
  {author} {\bibfnamefont {E.}~\bibnamefont {Pomjakushina}}, \bibinfo {author}
  {\bibfnamefont {P.}~\bibnamefont {{Qu\'emerais}}}, \ and\ \bibinfo {author}
  {\bibfnamefont {Z.}~\bibnamefont {Salman}},\ }\href {\doibase
  10.1103/PhysRevB.87.184427} {\bibfield  {journal} {\bibinfo  {journal} {Phys.
  Rev. B}\ }\textbf {\bibinfo {volume} {87}},\ \bibinfo {pages} {184427}
  (\bibinfo {year} {2013})}\BibitemShut {NoStop}%
\bibitem [{\citenamefont {Gardner}\ \emph {et~al.}(2011)\citenamefont
  {Gardner}, \citenamefont {Ehlers}, \citenamefont {Fouquet}, \citenamefont
  {Farago},\ and\ \citenamefont {Stewart}}]{Gardner11}%
  \BibitemOpen
  \bibfield  {author} {\bibinfo {author} {\bibfnamefont {J.~S.}\ \bibnamefont
  {Gardner}}, \bibinfo {author} {\bibfnamefont {G.}~\bibnamefont {Ehlers}},
  \bibinfo {author} {\bibfnamefont {P.}~\bibnamefont {Fouquet}}, \bibinfo
  {author} {\bibfnamefont {B.}~\bibnamefont {Farago}}, \ and\ \bibinfo {author}
  {\bibfnamefont {J.~R.}\ \bibnamefont {Stewart}},\ }\href
  {http://iopscience.iop.org/0953-8984/23/16/164220} {\bibfield  {journal}
  {\bibinfo  {journal} {J. Phys.: Condens. Matter}\ }\textbf {\bibinfo {volume}
  {23}},\ \bibinfo {pages} {164220} (\bibinfo {year} {2011})}\BibitemShut
  {NoStop}%
\bibitem [{\citenamefont {Bloembergen}(1949)}]{Bloembergen49}%
  \BibitemOpen
  \bibfield  {author} {\bibinfo {author} {\bibfnamefont {N.}~\bibnamefont
  {Bloembergen}},\ }\href {\doibase 10.1016/0031-8914(49)90114-7} {\bibfield
  {journal} {\bibinfo  {journal} {Physica}\ }\textbf {\bibinfo {volume} {15}},\
  \bibinfo {pages} {386 } (\bibinfo {year} {1949})}\BibitemShut {NoStop}%
\bibitem [{\citenamefont {Van~Hove}(1954)}]{vanHove54}%
  \BibitemOpen
  \bibfield  {author} {\bibinfo {author} {\bibfnamefont {L.}~\bibnamefont
  {Van~Hove}},\ }\href {\doibase 10.1103/PhysRev.95.1374} {\bibfield  {journal}
  {\bibinfo  {journal} {Phys. Rev.}\ }\textbf {\bibinfo {volume} {95}},\
  \bibinfo {pages} {1374} (\bibinfo {year} {1954})}\BibitemShut {NoStop}%
\bibitem [{\citenamefont {Benner}\ and\ \citenamefont
  {Boucher}(1990)}]{deJongh90}%
  \BibitemOpen
  \bibfield  {author} {\bibinfo {author} {\bibfnamefont {H.}~\bibnamefont
  {Benner}}\ and\ \bibinfo {author} {\bibfnamefont {J.~P.}\ \bibnamefont
  {Boucher}},\ }in\ \href@noop {} {\emph {\bibinfo {booktitle} {Magnetic
  properties of layered transition metal compounds}}},\ \bibinfo {editor}
  {edited by\ \bibinfo {editor} {\bibfnamefont {L.~J.}\ \bibnamefont
  {de~Jongh}}}\ (\bibinfo  {publisher} {Kluwer Academic Publishers},\ \bibinfo
  {year} {1990})\ pp.\ \bibinfo {pages} {323--378}\BibitemShut {NoStop}%
\bibitem [{\citenamefont {Sirker}\ \emph {et~al.}(2011)\citenamefont {Sirker},
  \citenamefont {Pereira},\ and\ \citenamefont {Affleck}}]{Sirker11}%
  \BibitemOpen
  \bibfield  {author} {\bibinfo {author} {\bibfnamefont {J.}~\bibnamefont
  {Sirker}}, \bibinfo {author} {\bibfnamefont {R.~G.}\ \bibnamefont {Pereira}},
  \ and\ \bibinfo {author} {\bibfnamefont {I.}~\bibnamefont {Affleck}},\ }\href
  {\doibase 10.1103/PhysRevB.83.035115} {\bibfield  {journal} {\bibinfo
  {journal} {Phys. Rev. B}\ }\textbf {\bibinfo {volume} {83}},\ \bibinfo
  {pages} {035115} (\bibinfo {year} {2011})}\BibitemShut {NoStop}%
\bibitem [{\citenamefont {Villain}(1979)}]{Villain79}%
  \BibitemOpen
  \bibfield  {author} {\bibinfo {author} {\bibfnamefont {J.}~\bibnamefont
  {Villain}},\ }\href@noop {} {\bibfield  {journal} {\bibinfo  {journal} {Z.
  Phys. B}\ }\textbf {\bibinfo {volume} {33}},\ \bibinfo {pages} {31} (\bibinfo
  {year} {1979})}\BibitemShut {NoStop}%
\bibitem [{\citenamefont {Hermele}\ \emph {et~al.}(2004)\citenamefont
  {Hermele}, \citenamefont {Fisher},\ and\ \citenamefont
  {Balents}}]{Hermele04}%
  \BibitemOpen
  \bibfield  {author} {\bibinfo {author} {\bibfnamefont {M.}~\bibnamefont
  {Hermele}}, \bibinfo {author} {\bibfnamefont {M.~P.~A.}\ \bibnamefont
  {Fisher}}, \ and\ \bibinfo {author} {\bibfnamefont {L.}~\bibnamefont
  {Balents}},\ }\href@noop {} {\bibfield  {journal} {\bibinfo  {journal} {Phys.
  Rev. B}\ }\textbf {\bibinfo {volume} {69}},\ \bibinfo {pages} {064404}
  (\bibinfo {year} {2004})}\BibitemShut {NoStop}%
\bibitem [{\citenamefont {Melko}\ and\ \citenamefont
  {Gingras}(2004)}]{Melko04}%
  \BibitemOpen
  \bibfield  {author} {\bibinfo {author} {\bibfnamefont {R.~G.}\ \bibnamefont
  {Melko}}\ and\ \bibinfo {author} {\bibfnamefont {M.~J.~P.}\ \bibnamefont
  {Gingras}},\ }\href {http://stacks.iop.org/0953-8984/16/i=43/a=R02}
  {\bibfield  {journal} {\bibinfo  {journal} {Journal of Physics: Condens.
  Matter}\ }\textbf {\bibinfo {volume} {16}},\ \bibinfo {pages} {R1277}
  (\bibinfo {year} {2004})}\BibitemShut {NoStop}%
\bibitem [{\citenamefont {Lee}\ \emph {et~al.}(2002)\citenamefont {Lee},
  \citenamefont {Broholm}, \citenamefont {Ratcliff}, \citenamefont
  {Gasparovic}, \citenamefont {Huang}, \citenamefont {Kim},\ and\ \citenamefont
  {Cheong}}]{Lee02}%
  \BibitemOpen
  \bibfield  {author} {\bibinfo {author} {\bibfnamefont {S.-H.}\ \bibnamefont
  {Lee}}, \bibinfo {author} {\bibfnamefont {C.}~\bibnamefont {Broholm}},
  \bibinfo {author} {\bibfnamefont {W.}~\bibnamefont {Ratcliff}}, \bibinfo
  {author} {\bibfnamefont {G.}~\bibnamefont {Gasparovic}}, \bibinfo {author}
  {\bibfnamefont {Q.}~\bibnamefont {Huang}}, \bibinfo {author} {\bibfnamefont
  {T.~H.}\ \bibnamefont {Kim}}, \ and\ \bibinfo {author} {\bibfnamefont
  {S.-W.}\ \bibnamefont {Cheong}},\ }\href {\doibase 10.1038/nature00964}
  {\bibfield  {journal} {\bibinfo  {journal} {Nature}\ }\textbf {\bibinfo
  {volume} {418}},\ \bibinfo {pages} {856} (\bibinfo {year}
  {2002})}\BibitemShut {NoStop}%
\bibitem [{\citenamefont {Chung}\ \emph {et~al.}(2005)\citenamefont {Chung},
  \citenamefont {Matsuda}, \citenamefont {Lee}, \citenamefont {Kakurai},
  \citenamefont {Ueda}, \citenamefont {Sato}, \citenamefont {Takagi},
  \citenamefont {Hong},\ and\ \citenamefont {Park}}]{Chung05}%
  \BibitemOpen
  \bibfield  {author} {\bibinfo {author} {\bibfnamefont {J.-H.}\ \bibnamefont
  {Chung}}, \bibinfo {author} {\bibfnamefont {M.}~\bibnamefont {Matsuda}},
  \bibinfo {author} {\bibfnamefont {S.-H.}\ \bibnamefont {Lee}}, \bibinfo
  {author} {\bibfnamefont {K.}~\bibnamefont {Kakurai}}, \bibinfo {author}
  {\bibfnamefont {H.}~\bibnamefont {Ueda}}, \bibinfo {author} {\bibfnamefont
  {T.~J.}\ \bibnamefont {Sato}}, \bibinfo {author} {\bibfnamefont
  {H.}~\bibnamefont {Takagi}}, \bibinfo {author} {\bibfnamefont {K.-P.}\
  \bibnamefont {Hong}}, \ and\ \bibinfo {author} {\bibfnamefont
  {S.}~\bibnamefont {Park}},\ }\href {\doibase 10.1103/PhysRevLett.95.247204}
  {\bibfield  {journal} {\bibinfo  {journal} {Phys. Rev. Lett.}\ }\textbf
  {\bibinfo {volume} {95}},\ \bibinfo {pages} {247204} (\bibinfo {year}
  {2005})}\BibitemShut {NoStop}%
\bibitem [{\citenamefont {Tomiyasu}\ \emph {et~al.}(2008)\citenamefont
  {Tomiyasu}, \citenamefont {Suzuki}, \citenamefont {Toki}, \citenamefont
  {Itoh}, \citenamefont {Matsuura}, \citenamefont {Aso},\ and\ \citenamefont
  {Yamada}}]{Tomiyasu08}%
  \BibitemOpen
  \bibfield  {author} {\bibinfo {author} {\bibfnamefont {K.}~\bibnamefont
  {Tomiyasu}}, \bibinfo {author} {\bibfnamefont {H.}~\bibnamefont {Suzuki}},
  \bibinfo {author} {\bibfnamefont {M.}~\bibnamefont {Toki}}, \bibinfo {author}
  {\bibfnamefont {S.}~\bibnamefont {Itoh}}, \bibinfo {author} {\bibfnamefont
  {M.}~\bibnamefont {Matsuura}}, \bibinfo {author} {\bibfnamefont
  {N.}~\bibnamefont {Aso}}, \ and\ \bibinfo {author} {\bibfnamefont
  {K.}~\bibnamefont {Yamada}},\ }\href {\doibase
  10.1103/PhysRevLett.101.177401} {\bibfield  {journal} {\bibinfo  {journal}
  {Phys. Rev. Lett.}\ }\textbf {\bibinfo {volume} {101}},\ \bibinfo {pages}
  {177401} (\bibinfo {year} {2008})}\BibitemShut {NoStop}%
\bibitem [{\citenamefont {Tomiyasu}\ \emph {et~al.}(2013)\citenamefont
  {Tomiyasu}, \citenamefont {Yokobori}, \citenamefont {Kousaka}, \citenamefont
  {Bewley}, \citenamefont {Guidi}, \citenamefont {Watanabe}, \citenamefont
  {Akimitsu},\ and\ \citenamefont {Yamada}}]{Tomiyasu13}%
  \BibitemOpen
  \bibfield  {author} {\bibinfo {author} {\bibfnamefont {K.}~\bibnamefont
  {Tomiyasu}}, \bibinfo {author} {\bibfnamefont {T.}~\bibnamefont {Yokobori}},
  \bibinfo {author} {\bibfnamefont {Y.}~\bibnamefont {Kousaka}}, \bibinfo
  {author} {\bibfnamefont {R.~I.}\ \bibnamefont {Bewley}}, \bibinfo {author}
  {\bibfnamefont {T.}~\bibnamefont {Guidi}}, \bibinfo {author} {\bibfnamefont
  {T.}~\bibnamefont {Watanabe}}, \bibinfo {author} {\bibfnamefont
  {J.}~\bibnamefont {Akimitsu}}, \ and\ \bibinfo {author} {\bibfnamefont
  {K.}~\bibnamefont {Yamada}},\ }\href {\doibase
  10.1103/PhysRevLett.110.077205} {\bibfield  {journal} {\bibinfo  {journal}
  {Phys. Rev. Lett.}\ }\textbf {\bibinfo {volume} {110}},\ \bibinfo {pages}
  {077205} (\bibinfo {year} {2013})}\BibitemShut {NoStop}%
\bibitem [{\citenamefont {Ballou}\ \emph {et~al.}(1996)\citenamefont {Ballou},
  \citenamefont {Leli\`evre-Berna},\ and\ \citenamefont {F\aa{}k}}]{Ballou96}%
  \BibitemOpen
  \bibfield  {author} {\bibinfo {author} {\bibfnamefont {R.}~\bibnamefont
  {Ballou}}, \bibinfo {author} {\bibfnamefont {E.}~\bibnamefont
  {Leli\`evre-Berna}}, \ and\ \bibinfo {author} {\bibfnamefont
  {B.}~\bibnamefont {F\aa{}k}},\ }\href {\doibase 10.1103/PhysRevLett.76.2125}
  {\bibfield  {journal} {\bibinfo  {journal} {Phys. Rev. Lett.}\ }\textbf
  {\bibinfo {volume} {76}},\ \bibinfo {pages} {2125} (\bibinfo {year}
  {1996})}\BibitemShut {NoStop}%
\bibitem [{\citenamefont {Yavors'kii}\ \emph {et~al.}(2008)\citenamefont
  {Yavors'kii}, \citenamefont {Fennell}, \citenamefont {Gingras},\ and\
  \citenamefont {Bramwell}}]{Yavors08}%
  \BibitemOpen
  \bibfield  {author} {\bibinfo {author} {\bibfnamefont {T.}~\bibnamefont
  {Yavors'kii}}, \bibinfo {author} {\bibfnamefont {T.}~\bibnamefont {Fennell}},
  \bibinfo {author} {\bibfnamefont {M.~J.~P.}\ \bibnamefont {Gingras}}, \ and\
  \bibinfo {author} {\bibfnamefont {S.~T.}\ \bibnamefont {Bramwell}},\
  }\href@noop {} {\bibfield  {journal} {\bibinfo  {journal} {Phys. Rev. Lett.}\
  }\textbf {\bibinfo {volume} {101}},\ \bibinfo {pages} {037204} (\bibinfo
  {year} {2008})}\BibitemShut {NoStop}%
\bibitem [{\citenamefont {Ryzhkin}(2005)}]{Ryzhkin05}%
  \BibitemOpen
  \bibfield  {author} {\bibinfo {author} {\bibfnamefont {I.~A.}\ \bibnamefont
  {Ryzhkin}},\ }\href@noop {} {\bibfield  {journal} {\bibinfo  {journal}
  {JETP}\ }\textbf {\bibinfo {volume} {101}},\ \bibinfo {pages} {481} (\bibinfo
  {year} {2005})}\BibitemShut {NoStop}%
\bibitem [{\citenamefont {Castelnovo}\ \emph {et~al.}(2008)\citenamefont
  {Castelnovo}, \citenamefont {Moessner},\ and\ \citenamefont
  {Sondhi}}]{Castelnovo08}%
  \BibitemOpen
  \bibfield  {author} {\bibinfo {author} {\bibfnamefont {C.}~\bibnamefont
  {Castelnovo}}, \bibinfo {author} {\bibfnamefont {R.}~\bibnamefont
  {Moessner}}, \ and\ \bibinfo {author} {\bibfnamefont {S.~L.}\ \bibnamefont
  {Sondhi}},\ }\href {\doibase 10.1038/nature06433} {\bibfield  {journal}
  {\bibinfo  {journal} {Nature}\ }\textbf {\bibinfo {volume} {451}},\ \bibinfo
  {pages} {42} (\bibinfo {year} {2008})}\BibitemShut {NoStop}%
\bibitem [{\citenamefont {Arima}(2013)}]{Arima13}%
  \BibitemOpen
  \bibfield  {author} {\bibinfo {author} {\bibfnamefont {T.-H.}\ \bibnamefont
  {Arima}},\ }\href {\doibase 10.7566/JPSJ.82.013705} {\bibfield  {journal}
  {\bibinfo  {journal} {J. Phys. Soc. Jpn.}\ }\textbf {\bibinfo {volume}
  {82}},\ \bibinfo {pages} {013705} (\bibinfo {year} {2013})}\BibitemShut
  {NoStop}%
\bibitem [{Note2()}]{Note2}%
  \BibitemOpen
  \bibinfo {note} {See, e.g. Refs.~\protect \rev@citealpnum
  {Zheng05,Yaouanc05a} for the lattice of corner sharing tetrahedra,
  Ref.~\protect \rev@citealpnum {Mendels07} for the kagome lattice and Ref.~
  \protect \rev@citealpnum {Dalmas12} for the triangular lattice}\BibitemShut
  {NoStop}%
\bibitem [{\citenamefont {Mendels}\ \emph {et~al.}(2007)\citenamefont
  {Mendels}, \citenamefont {Bert}, \citenamefont {de~Vries}, \citenamefont
  {Olariu}, \citenamefont {Harrison}, \citenamefont {Duc}, \citenamefont
  {Trombe}, \citenamefont {Lord}, \citenamefont {Amato},\ and\ \citenamefont
  {Baines}}]{Mendels07}%
  \BibitemOpen
  \bibfield  {author} {\bibinfo {author} {\bibfnamefont {P.}~\bibnamefont
  {Mendels}}, \bibinfo {author} {\bibfnamefont {F.}~\bibnamefont {Bert}},
  \bibinfo {author} {\bibfnamefont {M.~A.}\ \bibnamefont {de~Vries}}, \bibinfo
  {author} {\bibfnamefont {A.}~\bibnamefont {Olariu}}, \bibinfo {author}
  {\bibfnamefont {A.}~\bibnamefont {Harrison}}, \bibinfo {author}
  {\bibfnamefont {F.}~\bibnamefont {Duc}}, \bibinfo {author} {\bibfnamefont
  {J.~C.}\ \bibnamefont {Trombe}}, \bibinfo {author} {\bibfnamefont {J.~S.}\
  \bibnamefont {Lord}}, \bibinfo {author} {\bibfnamefont {A.}~\bibnamefont
  {Amato}}, \ and\ \bibinfo {author} {\bibfnamefont {C.}~\bibnamefont
  {Baines}},\ }\href {\doibase 10.1103/PhysRevLett.98.077204} {\bibfield
  {journal} {\bibinfo  {journal} {Phys. Rev. Lett.}\ }\textbf {\bibinfo
  {volume} {98}},\ \bibinfo {pages} {077204} (\bibinfo {year}
  {2007})}\BibitemShut {NoStop}%
\bibitem [{\citenamefont {Dalmas~de R\'eotier}\ \emph
  {et~al.}(2012{\natexlab{b}})\citenamefont {Dalmas~de R\'eotier},
  \citenamefont {Yaouanc}, \citenamefont {MacLaughlin}, \citenamefont {Zhao},
  \citenamefont {Higo}, \citenamefont {Nakatsuji}, \citenamefont {Nambu},
  \citenamefont {Marin}, \citenamefont {Lapertot}, \citenamefont {Amato},\ and\
  \citenamefont {Baines}}]{Dalmas12}%
  \BibitemOpen
  \bibfield  {author} {\bibinfo {author} {\bibfnamefont {P.}~\bibnamefont
  {Dalmas~de R\'eotier}}, \bibinfo {author} {\bibfnamefont {A.}~\bibnamefont
  {Yaouanc}}, \bibinfo {author} {\bibfnamefont {D.~E.}\ \bibnamefont
  {MacLaughlin}}, \bibinfo {author} {\bibfnamefont {S.}~\bibnamefont {Zhao}},
  \bibinfo {author} {\bibfnamefont {T.}~\bibnamefont {Higo}}, \bibinfo {author}
  {\bibfnamefont {S.}~\bibnamefont {Nakatsuji}}, \bibinfo {author}
  {\bibfnamefont {Y.}~\bibnamefont {Nambu}}, \bibinfo {author} {\bibfnamefont
  {C.}~\bibnamefont {Marin}}, \bibinfo {author} {\bibfnamefont
  {G.}~\bibnamefont {Lapertot}}, \bibinfo {author} {\bibfnamefont
  {A.}~\bibnamefont {Amato}}, \ and\ \bibinfo {author} {\bibfnamefont
  {C.}~\bibnamefont {Baines}},\ }\href {\doibase 10.1103/PhysRevB.85.140407}
  {\bibfield  {journal} {\bibinfo  {journal} {Phys. Rev. B}\ }\textbf {\bibinfo
  {volume} {85}},\ \bibinfo {pages} {140407} (\bibinfo {year}
  {2012}{\natexlab{b}})}\BibitemShut {NoStop}%
\end{thebibliography}%

\end{document}